\documentclass[aps,pra,twocolumn,superscriptaddress,longbibliography,preprintnumbers]{revtex4-2} 

\usepackage{amsmath}
\usepackage{amsfonts}
\usepackage{amssymb}
\usepackage{graphicx}
\usepackage{subfigure}
\usepackage{color}
\usepackage{accents}
\usepackage[colorlinks=true, citecolor=cyan]{hyperref}
\usepackage{enumitem}

\newcommand{\Tr}{\mbox{Tr}}

\newcommand{\ket}[1]{|#1\rangle}
\newcommand{\bra}[1]{\langle #1|}

\makeatletter 
\newsavebox{\@brx}
\newcommand{\llangle}[1][]{\savebox{\@brx}{\(\m@th{#1\langle}\)}%
  \mathopen{\copy\@brx\kern-0.5\wd\@brx\usebox{\@brx}}}
\newcommand{\rrangle}[1][]{\savebox{\@brx}{\(\m@th{#1\rangle}\)}%
  \mathclose{\copy\@brx\kern-0.5\wd\@brx\usebox{\@brx}}}
\makeatother

\newlength{\dhatheight} 

\newtheorem{result}{Result}
\newtheorem{corollary}[result]{Corollary}
\newtheorem{conjecture}[result]{Conjecture}

\newcommand{\qed}{\nobreak \ifvmode \relax \else
      \ifdim\lastskip<1.5em \hskip-\lastskip
      \hskip1.5em plus0em minus0.5em \fi \nobreak
      \vrule height0.75em width0.5em depth0.25em\fi}

\begin{document}

\title{Information transfer along the causal lightcone of a brickwork quantum circuit}
\author{Shivansh Singh}
\email[]{singhs13@tcd.ie}
\affiliation{School of Theoretical Physics, Dublin Institute for Advanced Studies, 10 Burlington Road, Dublin 4, Ireland.}
\affiliation{School of Physics, Trinity College Dublin, College Green, Dublin 2, Ireland}
\author{Shane Dooley}
\email[]{dooleysh@gmail.com}
\affiliation{School of Theoretical Physics, Dublin Institute for Advanced Studies, 10 Burlington Road, Dublin 4, Ireland.}
\date{\today}

\begin{abstract}
Understanding how local information propagates through many-body quantum systems is a central problem in nonequilibrium dynamics, with important implications for quantum communication, state transfer, and remote sensing. In this work, we investigate information transfer along a one-dimensional open chain of qudits, focusing on the task of recovering information initially encoded at one end via measurements performed at the opposite end. By restricting the dynamics to brickwork quantum circuits, and considering small $M$-qudit subsystems on the causal ``lightcone'' of the circuit, we obtain several results valid even for large system sizes $N$ or for nonintegrable global dynamics. Within this framework, lossless information transfer is linked to the existence of peripheral eigenvalues of a quantum channel $\Phi_M$, which governs the evolution of the $M$-qudit local subsystem along the lightcone. We investigate conditions under which brickwork circuits admit such peripheral eigenvalues. For qubit chains and $M=1$, we show that the dual-unitary property is necessary, whereas for larger local subsystems ($M \geq 2$) or higher-dimensional qudits, this requirement can be relaxed. Perhaps surprisingly, we can use the peripheral eigenvalue condition to construct examples exhibiting lossless information transfer through chains of arbitrary system size $N$, even when the underlying circuit dynamics is nonintegrable and thermalising at long times.
\end{abstract}


\maketitle

\section{Introduction}

A key question in nonequilibrium many-body physics concerns how information initially localized in a small region spreads under quantum dynamics, characterising the decoherence and thermalisation of the system \cite{Rig-08,DAl-16,Mor-18,Aba-19}. Beyond its fundamental interest, this process plays an important role in quantum state transfer \cite{Bos-03a,Chr-04a,Kay-10} and remote quantum sensing \cite{Kiu-17a, Jon-20a, Mon-22a}, where one tries to engineer systems in which the information is transmitted faithfully between distant locations \cite{Li-18a,Xia-24a,Doo-25a}.

\begin{figure} 
    \centering
    \includegraphics[width=\columnwidth]{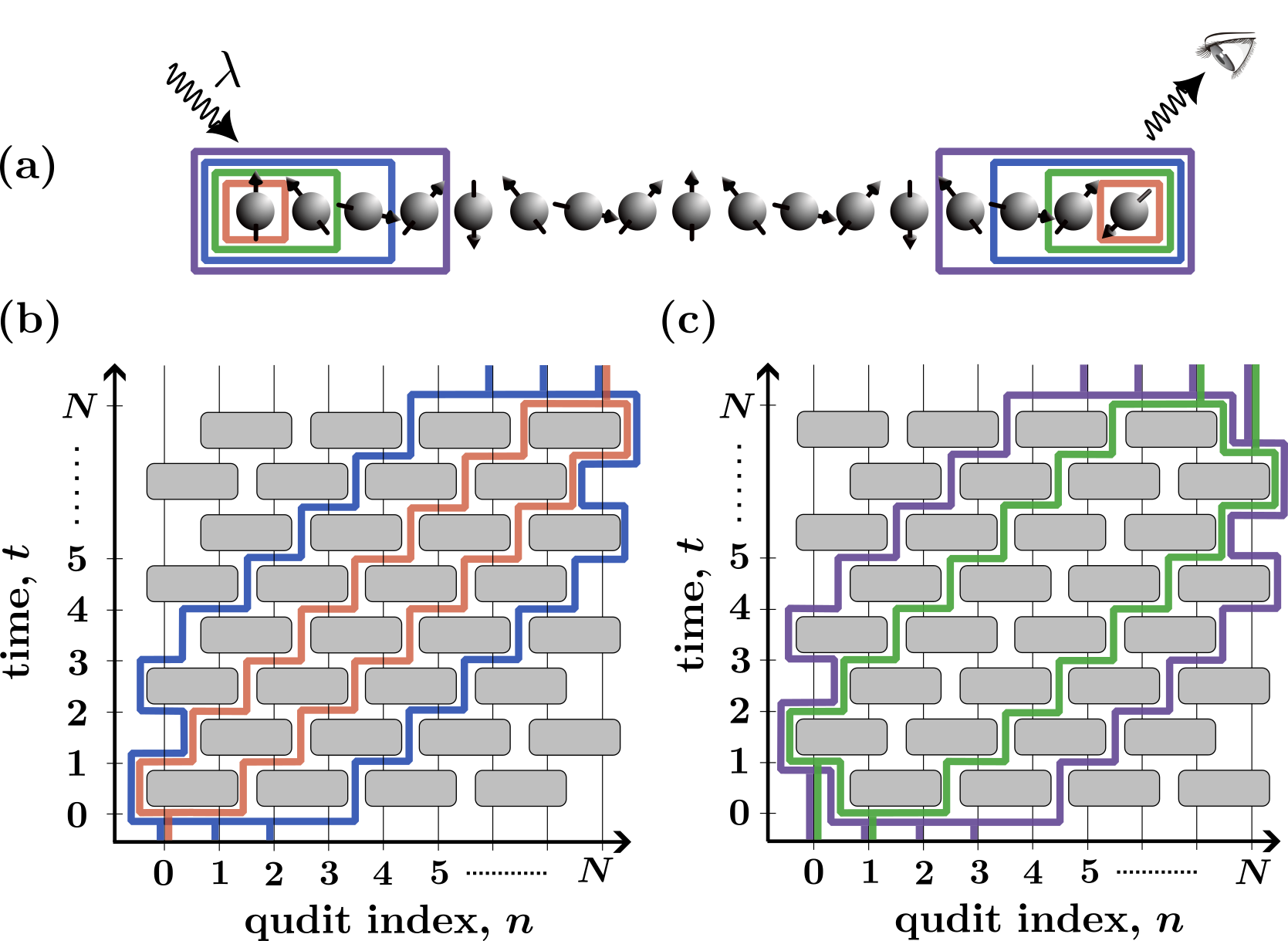}
    
    \caption{(a) We consider the problem of information transfer through a thermal Floquet chain of $N+1$ qudits. Some information $\lambda$ is initially encoded in the state of the first $M$ qudits (the red, blue, green, purple boxes, for $M=1,2,3,4$, respectively), and estimated through measurement of the last $M$ qudits at a later time. (b,c) We restrict to the class of Floquet models that can be mapped to brickwork circuits composed of two-qudit unitary channels $\hat{\mathcal{U}}_2 = \hat{U} \cdot \hat{U}^\dagger$ (gray boxes). Given the initial state $\hat{\rho}^{(\lambda)}(0) = \hat{\rho}_{{\rm in},M}^{(\lambda)} \otimes (\hat{\mathbb{I}}/d)^{\otimes (N-M+1)}$ for small $M$, we can efficiently compute the reduced density matrix $\hat{\rho}^{(\lambda)}_{{\rm out},M}(t)$ of the last $M$ qudits at the later time $t=N$ on the ``lightcone'', which allows us study the information transfer problem even if the chain is long $N \gg 1$ or the model is non-integrable.}
    \label{fig:schematic}
\end{figure}

In this work, we study the transfer of information through a one-dimensional chain of qudits, focusing on the task of recovering information initially encoded at one end of the chain via measurements performed at the opposite end [see Fig.~\ref{fig:schematic}(a)]. On the one hand, in generic interacting systems this is a hopeless task, since initially local information becomes scrambled over time \cite{Rig-08,DAl-16,Mor-18,Mi-21a}, inhibiting the retrieval of the information at distant locations. On the other hand, finding conditions for lossless information transfer is often challenging, even numerically, as the dimension of the Hilbert space grows exponentially with system size, and the complexity of the time-evolved quantum state rapidly becomes intractable. 

We overcome the latter computational obstacles by focusing on a class of many-body Floquet systems that can be mapped to brickwork quantum circuits [see Fig.~\ref{fig:schematic}(b,c)]. These minimal models have emerged recently as a powerful setting for studying quantum dynamics, as they can admit exact solutions even in models that are nonintegrable \cite{Ber-18a, Pir-20a, Che-22a, Cla-21a, Log-24a, Fis-23b, Ber-26a, Fis-26a}. For example, their structure enables an exact and efficient description of dynamics along the causal ``lightcone'' of the circuit \cite{Ber-19a, Cla-20a, Kos-23a}. We leverage this partial solvability to investigate conditions under which the circuit supports lossless information transfer through the qudit chain.


Our paper is structured as follows. In Sec.~\ref{sec:info_transfer}, we introduce the system setup: information encoded in the initial state of the first $M$ qudits propagates through the chain and is recovered via measurements on the last $M$ qudits [see Fig.~\ref{fig:schematic}(a)]. We also define the figures of merit used to quantify information transfer, namely the trace distance for discretely encoded information and the quantum Fisher information for a continuous parameter encoding. 
In Sec.~\ref{sec:lightcone_dynamics}, we show that (provided $M$ is not too large) the reduced density matrix of the $M$ qudits can be computed efficiently along the lightcone of a brickwork circuit via repeated application of an $M$-qudit quantum channel $\Phi_M$. We note that the channel $\Phi_M$ was also studied in Ref. \cite{Cla-22a} for a special integrable example, and in Ref. \cite{Ber-19a} for $M=1$. 
In Sec.~\ref{sec:classes}, we demonstrate that lossless information transfer along the lightcone occurs if and only if the channel $\Phi_M$ possesses peripheral eigenvalues, a result recently reported in Refs. \cite{Faw-25a,Sin-25a,Sin-26}, and closely related to the possibility of solitons in the circuit \cite{Ber-20c}.
In Sec.~\ref{sec:generic_loss}, we verify that typical brickwork circuits, which are nonintegrable and thermalising, do not have peripheral eigenvalues, leading to information loss. Guided by the peripheral eigenvalue criterion, in Sec.~\ref{sec:lossless_info_transfer} we identify classes of circuits that enable lossless transfer. In particular, for $M=1$ in qubit chains (Sec.~\ref{sec:d2_M1}), we show that lossless transfer can occur only in dual-unitary circuits. In contrast, for $M>1$ in qubit chains (Sec.~\ref{sec:d2_M2}) and for qutrit chains (Sec.~\ref{sec:d3_M1}), lossless transfer can arise even in non-dual-unitary circuits. 
Finally, in Sec.~\ref{sec:example}, we present an explicit example of a brickwork circuit that shows that lossless information transfer can occur even in models that are non-integrable and appear to satisfy the eigenstate thermalisation hypothesis (ETH) — a regime associated with local thermalisation and information scrambling at long evolution times \cite{Rig-08,DAl-16,Mor-18,Mi-21a}.

Taken together, our results reveal the usefulness of the peripheral eigenvalue criterion to determine when information can be transmitted without loss along the causal lightcone of a circuit. More broadly, they highlight the power of minimal, exactly solvable circuit models as a framework for obtaining insights into quantum information dynamics beyond integrable settings.

\section{Information transfer through a thermal Floquet qudit chain}
\label{sec:info_transfer}

Consider a chain of $N+1$ qudits (each with a $d$-dimensional local Hilbert space), labelled $n\in\{0, 1, 2, \hdots, N \}$, and assume for convenience that $N$ is even (i.e., the number of qudits $N+1$ is odd). The qudits evolve by a time-dependent local Hamiltonian $\hat{\mathbb{H}}(t) = \hat{\mathbb{H}}(t + \tau)$ with a period $\tau$. We focus on the stroboscopic, discrete-time dynamics, given by repeated application of the Floquet unitary: \begin{equation} \hat{\mathbb{U}} = \mathcal{T} \exp \Big[ -\frac{i}{\hbar}\int_0^\tau dt \hat{\mathbb{H}}(t) \Big] , \end{equation} where $\mathcal{T}$ is the time-ordering operator. It follows that time evolution for $k\in\mathbb{Z}$ Floquet periods (i.e., for a discrete time $t=k\tau$) is implemented by the unitary operator $\hat{\mathbb{U}}^k$ (the Floquet unitary taken to the power of $k$). We note that, since continuous time-translation symmetry is broken in the Hamiltonian, energy is not a conserved quantity and -- in the absence of any other local conserved quantities -- the local thermal state for each local subsystem of $M \ll N$ qudits is given by its infinite temperature density matrix $(\hat{\mathbb{I}}_d / d)^{\otimes M}$, where $\hat{\mathbb{I}}_d$ is the single-qudit identity operator. 

We consider initial states of the form: \begin{equation} \hat{\rho} (0) = \hat{\rho}_{{\rm in}, M} (0) \otimes \left( \frac{\hat{\mathbb{I}}_d}{d} \right)^{\otimes (N-M+1)} , \label{eq:rho_0} \end{equation} where $\hat{\rho}_{{\rm in}, M} (0) = {\rm Tr}_{M,...,N} [\hat{\rho}(0)]$ is a density matrix for the first $M$ qudits at time $t=0$, and the remaining qudits are in thermal equilibrium (their maximally mixed state). We would like to understand how information encoded in the initial state $\hat{\rho}_{{\rm in}, M} (0)$ of the first $M$-qudits is propagated through the thermal chain and can be recovered by a measurement of the reduced density matrix $\hat{\rho}_{{\rm out}, M} (t) \equiv {\rm Tr}_{0,...,N-M} [\hat{\rho}(t)]$ of the last $M$ qudits at a later time $t$ [see Fig. \ref{fig:schematic}(a)]. This can be quantified in different ways, depending on whether the encoding of the information in the initial state of the first $M$ qudits is discrete or continuous.

Discrete information can, for example, be encoded in the initial state of the first $M$ qudits by preparing one of two states, $\hat{\rho}_{{\rm in}, M}^{(1)} (0)$ or $\hat{\rho}_{{\rm in}, M}^{(2)} (0)$ which, after evolution for time $t$, gives either the output state $\hat{\rho}_{{\rm out}, M}^{(1)} (t)$ or $\hat{\rho}_{{\rm out}, M}^{(2)} (t)$ of the last $M$ qudits. The ability to distinguish between two quantum states $\hat{\rho}^{(1)}$ and $\hat{\rho}^{(2)}$ by any measurement is quantified by the trace distance $D [\hat{\rho}^{(1)}, \hat{\rho}^{(2)} ]$, which is defined as \cite{Nie-00}: 
\begin{equation} D [\hat{X}, \hat{Y} ] = \frac{1}{2} \Tr \sqrt{ (\hat{X} - \hat{Y})^\dagger (\hat{X} - \hat{Y}) } , \label{eq:trace_distance} \end{equation} for arbitrary operators $\hat{X}$ and $\hat{Y}$. We can therefore quantify the loss of information as a result of transmission through the chain with the ratio:
\begin{equation} \eta_D (t,N,M) = \frac{D[ \hat{\rho}^{(1)}_{{\rm out},M}(t), \hat{\rho}^{(2)}_{{\rm out},M}(t) ]}{D [ \hat{\rho}_{{\rm in}, M}^{(1)}(0), \hat{\rho}_{{\rm in},M}^{(2)}(0) ]} . \end{equation} Since the trace distance cannot increase under any quantum channel \cite{Nie-00}, we must have $\eta_D \leq 1$, with $\eta_D = 1$ if the information is transmitted faithfully through the chain without any loss. 

Alternatively, continuous information can be encoded in a variable $\lambda \in \mathbb{R}$, which parameterises the initial state $\hat{\rho}_{{\rm in},M}^{(\lambda)}(0)$ of the first $M$ qudits. After evolution for a time $t$, the last $M$ qudits are in the state $\hat{\rho}^{(\lambda)}_{{\rm out},M}(t)$. The ability to estimate a parameter $\lambda$ encoded in a quantum state $\hat{\rho}^{(\lambda)}$ is quantified by the quantum Fisher information (QFI) \cite{Bra-94,Par-09}:
\begin{equation} F[\hat{\rho}^{(\lambda)}] = 2 \sum_{\substack{i,j \\ p_i + p_j \neq 0}} \frac{| \bra{\psi_i} \partial_\lambda \hat{\rho}^{(\lambda)} \ket{\psi_j} |^2}{p_i + p_j} , \label{eq:QFI} \end{equation} where $\hat{\rho}^{(\lambda)} = \sum_i p_i \ket{\psi_i}\bra{\psi_i}$ is the spectral decomposition of the parameterised density matrix. In this case, we can measure the loss of information as a result of transmission through the chain with the ratio:
\begin{equation} \eta_F (t, N, M) =
\sqrt{\frac{F[\hat{\rho}^{(\lambda)}_{{\rm out},M} (t)]}{F[\hat{\rho}^{(\lambda)}_{{\rm in},M}(0) ]}} . \label{eq:etaF} \end{equation} Since the QFI cannot be amplified by passing the state through a quantum channel that is independent of the parameter $\lambda$ \cite{Nie-00}, we must have $\eta_F \leq 1$. Although it is possible to achieve the maximum value $\eta_F = 1$ (or $\eta_D = 1$ in the case of discrete encoding of information), typically information is lost during transmission through the chain resulting in $\eta_F < 1$ (or $\eta_D < 1$) as we will see later. 

\section{Dynamics along the ``lightcone'' of a brickwork circuit}
\label{sec:lightcone_dynamics}

Computing the time-evolved state $\hat{\rho}_{{\rm out},M} (t)$ (and from this, the information loss $\eta_D$ or $\eta_F$) is typically prohibitively difficult, even numerically. This is because it usually involves calculating the time-evolved state of the entire $(N+1)$-qudit system, before taking the partial trace to obtain the reduced density matrix of the last $M$ qudits. However, when $M$ is not too large, we can efficiently compute $\hat{\rho}_{{\rm out},M} (t)$ for a class of ``minimal'' local models for which the Floquet unitary can be expressed as a brickwork circuit, at special times $t=N$ along the ``lightcone'' of the brickwork circuit. This was first shown for $M=1$ in Ref. \cite{Ber-19a}.

More precisely, we assume that the Floquet unitary has the brickwork circuit structure $\hat{\mathbb{U}} =  \hat{\mathbb{U}}_o \hat{\mathbb{U}}_e$ (if $M$ is odd) or $\hat{\mathbb{U}} =  \hat{\mathbb{U}}_e \hat{\mathbb{U}}_o$ (if $M$ is even). Here, $\hat{\mathbb{U}}_e = \hat{U}^{\otimes N} \otimes \hat{\mathbb{I}}_d$ is called the ``even'' layer of two-qudit gates and $\hat{\mathbb{U}}_o = \hat{\mathbb{I}}_d \otimes \hat{U}^{\otimes N}$ is called the ``odd'' layer [see Fig. \ref{fig:schematic}(b, c)]. We also assume that each layer of odd or even gates takes half a Floquet period $\tau/2$ (which is consistent with the Floquet unitary taking a full period $\tau$) and we choose units of time such that $\tau = 2$ [see Fig. \ref{fig:schematic}(b, c)]. An example of a Floquet model which can be expressed as a brickwork circuit in this way is the kicked Ising model \cite{Ber-18a,Gut-20a,Kim-23a,Fis-26a}.  


To see how the brickwork structure enables efficient calculation of the state $\hat{\rho}_{{\rm out},M} (t)$ along the lightcone $t=N$, let us first consider $M=1$. We define the two-qudit unitary quantum channel $\mathcal{U}_2 [\hat{\rho}] \equiv \hat{U} \hat{\rho} \hat{U}^\dagger$, which conjugates an input two-qudit density matrix $\hat{\rho}$ with the two-qudit unitary $\hat{U}$. We can represent the channel pictorially as: \begin{equation} \mathcal{U}_2 = \begin{gathered} \includegraphics[scale=0.5]{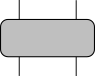} \end{gathered} \label{eq:U_2} , \end{equation} where each wire carries a one-qudit operator and time runs from bottom to top. Using this, we define the $M=1$ non-unitary quantum channel: 
\begin{equation} \Phi_1 [\hat{\rho}] \equiv \Tr_{q_1} \{ \mathcal{U}_2 [\hat{\rho}\otimes \hat{\mathbb{I}}_d/d] \} , \label{eq:Phi_1} \end{equation} where $\Tr_{q_1}$ denotes a partial trace of the first qudit in the tensor product. The channel $\Phi_1$ takes a one-qudit density matrix $\hat{\rho}$ as input, and propagates it one step forward in both time and space, giving the one-qudit reduced density matrix $\Phi_1 [\hat{\rho}]$ as output. We represent the channel pictorially as:
\begin{equation} \Phi_1 = \begin{gathered} \includegraphics[scale=0.5]{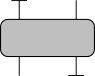} \end{gathered} \end{equation}
where the blocked wire on the bottom denotes an input maximally mixed state $\hat{\mathbb{I}}_d/d$ and the blocked wire on the top denotes the partial trace $\Tr_{q_1}$ of the first qudit.
The reduced density matrix $\hat{\rho}_{{\rm out},1} (t)$ of the last qudit at time $t=N$ is obtained by applying the quantum channel $\Phi_1$ repeatedly, $N$ times:
\begin{equation} \hat{\rho}_{{\rm out},1}(N) = \Phi_1^{N} [\hat{\rho}_{{\rm in},1}(0)] \sim \begin{gathered} \includegraphics[scale=0.5]{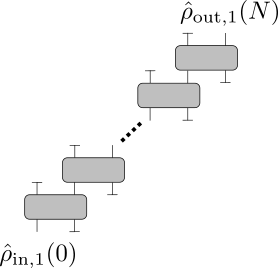} \end{gathered} , \label{eq:rho_N} \end{equation} as shown in the red region in Fig. \ref{fig:schematic}(b). 

Similarly, we define the three-qudit unitary quantum channel: \begin{eqnarray} \mathcal{U}_3 \equiv (\mathcal{U}_2 \otimes \mathcal{I}) ( \mathcal{I} \otimes \mathcal{U}_2 ) = \begin{gathered} \includegraphics[scale=0.5]{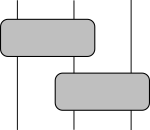} \end{gathered} \label{eq:U_3} \end{eqnarray} where $\mathcal{I}$ is the one-qudit identity channel. We use this to define the $M=2$ non-unitary channel:
\begin{equation} \Phi_2 [\hat{\rho}] \equiv \Tr_{q_1} \{ \mathcal{U}_3 [\hat{\rho}\otimes \hat{\mathbb{I}}_d/d] \} \sim \begin{gathered} \includegraphics[scale=0.5]{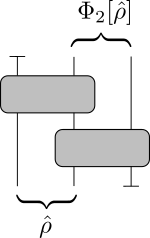} \end{gathered} \label{eq:Phi_2} \end{equation} which takes a two-qudit density matrix $\hat{\rho}$ as input and gives another two-qudit density $\Phi_2 [\hat{\rho}]$ matrix as output. The reduced density matrix $\hat{\rho}_{{\rm out},2}(t)$ of the last two qudits at time $t = N$ is obtained as: \begin{equation} \hat{\rho}_{{\rm out},2}(N) = \Phi_2^{N-1} [\hat{\rho}_{{\rm in},2}(0)] \sim \begin{gathered} \includegraphics[scale=0.5]{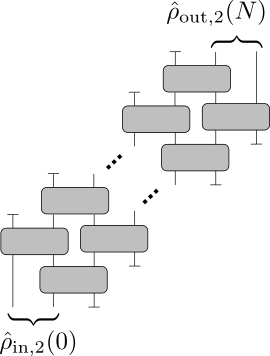} \end{gathered} \label{eq:rho_N-1,N} \end{equation} as shown in the green region of Fig. \ref{fig:schematic}(c).

We can generalise the unitary quantum channels in Eqs. \ref{eq:U_2} and \ref{eq:U_3} to $M \geq 2$ by recursively building on the corresponding channel for the previous value $M$:
\begin{eqnarray} \mathcal{U}_{M+1} \equiv (\mathcal{U}_2 \otimes \mathcal{I}^{\otimes (M-1)}) ( \mathcal{I} \otimes \mathcal{U}_{M} ) = \begin{gathered} \includegraphics[scale=0.5]{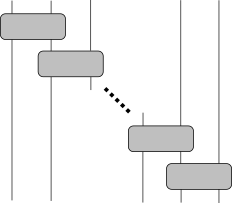} \end{gathered} . \end{eqnarray}
which we use to define the non-unitary quantum channel:
\begin{equation} \Phi_M [\hat{\rho}] \equiv \Tr_{q_1} \{ \mathcal{U}_{M+1} [\hat{\rho} \otimes \hat{\mathbb{I}}_d / d] \} \sim \begin{gathered} \includegraphics[scale=0.5]{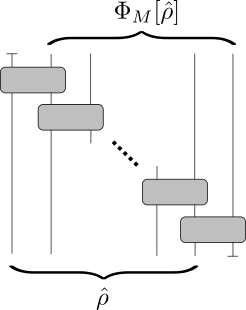} \end{gathered} \label{eq:Phi_M} \end{equation} which takes an $M$-qudit density matrix $\hat{\rho}$ as input and gives another $M$-qudit density matrix as output, generalising the quantum channels in Eqs. \ref{eq:Phi_1} and \ref{eq:Phi_2}.
The reduced density matrix $\hat{\rho}_{{\rm out},M}(t)$ of the last $M$ qudits at time $t = N$ is obtained as:
\begin{eqnarray} \hat{\rho}_{{\rm out},M}(N) &=& \Lambda_M^{\rm out} [\Phi_M^{N-M+1} [ \Lambda_M^{\rm in} [\hat{\rho}_{{\rm in},M} (0)]]] \nonumber \\ & & \nonumber \\ &\sim& \quad \begin{gathered} \includegraphics[scale=0.5]{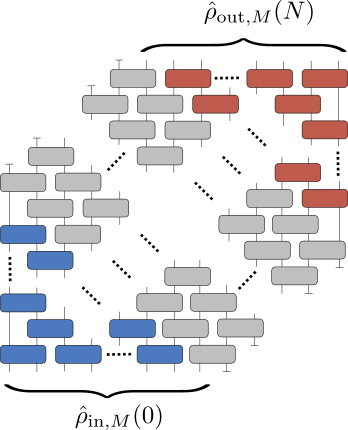} \end{gathered} \end{eqnarray}
where the pictorial representation above assumes that $M$ is odd, and we have introduced the $M$-qudit unitary channels $\Lambda_M^{\rm in}$ and $\Lambda_M^{\rm out}$ which are represented pictorially as ($M$ odd):
\begin{equation} \Lambda_M^{\rm in} = \begin{gathered} \includegraphics[scale=0.5]{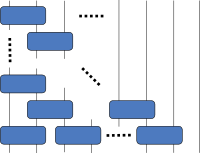} \end{gathered} \quad  \Lambda_M^{\rm out} = \begin{gathered} \includegraphics[scale=0.5]{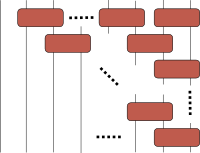} \end{gathered} \end{equation}
The precise definitions of the unitary channels $\Lambda_M^{\rm in/out}$ are given in Appendix \ref{app:a}. However, from the perspective of information transfer through the chain, we can neglect $\Lambda_M^{\rm in}$ and $\Lambda_M^{\rm out}$, since $\Lambda_M^{\rm in}$ can be absorbed into a redefinition of the initial state by the information-preserving unitary transformation $\hat{\rho}'_{{\rm in},M}(0) = \Lambda_M^{\rm in} [\hat{\rho}_{{\rm in},M} (0)]$ and $\Lambda_M^{\rm out}$ can be absorbed into a redefinition of the final state by the information-preserving unitary transformation $\hat{\rho}'_{{\rm out},M}(N) = (\Lambda_M^{\rm out})^{-1} [\hat{\rho}_{{\rm out},M}(N)]$. This leaves the repeated application of the non-unitary channel $\Phi_M$ as the important component of the dynamics through the chain of qudits:
\begin{eqnarray} \hat{\rho}'_{{\rm out},M}(N) = \Phi_M^{N-M+1}  [\hat{\rho}'_{{\rm in},M} (0)] . \label{eq:Phi_M_lightcone} \end{eqnarray} 
Since $\Phi_M$ is a quantum channel that takes as its input an $M$-qudit density operator (a $d^M\times d^M$ matrix, which can be vectorized as a $d^{2M}$-dimensional vector), and returns as its output another $M$-qudit density matrix, it can be represented as a $d^{2M} \times d^{2M}$ matrix, which grows exponentially with $M$ but is independent of $N$. This means that the action of the quantum channel $\Phi_M$ on its input state can be efficiently calculated for small $M$, which implies that the output reduced density matrix $\hat{\rho}_{{\rm out},M}(t)$ can be computed efficiently on the lightcone $t = N$, independent of the length $N$ of the chain, and independent of whether the brickwork circuit is integrable or non-integrable.

However, we emphasise that it is still generally intractable to compute the $M$-qudit reduced density matrix $\hat{\rho}_{{\rm out},M}(t)$ \emph{inside} the lightcone, i.e., for times $t > N$. An exception to this is when $M=1$ and the gates $\hat{U}$ have a special property called dual-unitarity, which means it is unitary not only as a time-evolution propagator, but also when it is interpreted as a propagator in the spatial direction \cite{Ber-26a}. In that case, it was shown in Ref. \cite{Ber-19a} that the single-qudit reduced density matrices inside the lightcone can be computed analytically, and are simply the single-qudit maximally mixed states.

Since the quantum channel $\Phi_M$ is the key object that determines the propagation of information along the lightcone of the chain, to further understand the different kinds of dynamics that are possible we should investigate properties of $\Phi_M$, many of which have been reported in previous works \cite{Ber-19a, Per-06a}. First, it can be easily verified from Eq. \ref{eq:Phi_M} that $\Phi_M [\hat{\mathbb{I}}_d^{\otimes M}] = \hat{\mathbb{I}}_d^{\otimes M}$, i.e., the quantum channel $\Phi_M$ is unital. We will refer to the identity eigenoperator $\hat{\mathbb{I}}_d^{\otimes M}$ and its unit eigenvalue as the trivial eigenoperator/eigenvalue of $\Phi_M$. Let us denote the \emph{non-trivial} eigenvalues and eigenoperators of $\Phi_M$ as: \begin{eqnarray} \Phi_M [\hat{A}_{\mu}] = z_\mu \hat{A}_\mu . \label{eq:Phi_eigs} \end{eqnarray} The unital property implies that the channel is contractive, i.e., its eigenvalues $z_\mu$ must all lie on the unit disc in the complex plane, $z_\mu \in \mathbb{C}$, $|z_\mu| \leq 1$ \cite{Per-06a, Ber-19a}. Moreover, one can show (see Appendix \ref{app:b}) that the non-trivial eigenoperators $\hat{A}_\mu$ can always be chosen to be traceless and that for each eigenvalue/eigenoperator $\Phi_M [\hat{A}_\mu] = z_\mu \hat{A}_\mu$ we also have $\Phi_M [\hat{A}_\mu^\dagger] = z_\mu^* \hat{A}_\mu^\dagger$. Finally, it is easy to see that if $\Phi_M [\hat{A}_{\mu}] = z_\mu \hat{A}_\mu$ are eigenoperators/eigenvalues of the channel $\Phi_M$ then $\Phi_{M+1} [\hat{A}_{\mu} \otimes \hat{\mathbb{I}}_d] = z_\mu \hat{A}_\mu \otimes \hat{\mathbb{I}}_d$ are eigenoperators/eigenvalues of the channel $\Phi_{M+1}$, i.e., the spectrum of $\Phi_M$ is a subset of the spectrum of $\Phi_{M+1}$ \footnote{A consequence is that, if for a given two-qudit unitary $\hat{U}$ the corresponding quantum channel $\Phi_M (\hat{U})$ is non-unitary, then $\Phi_{M'}(\hat{U})$ is also non-unitary for any $M' > M$, including in the limit $M' \to \infty$ where one might naively expect the channel to become increasingly unitary.}.

Let us assume that the quantum channel $\Phi_M$ is diagonalisable, so that any initial qudit state $\hat{\rho}'_{0,\hdots,M-1} (0)$ can be expressed as a linear combination of its eigenoperators:
\begin{equation} \hat{\rho}'_{{\rm in},M} (0) = \frac{1}{d^M} \Big\{ \hat{\mathbb{I}}_d^{\otimes M} + \sum_\mu [ f_\mu \hat{A}_\mu + f_\mu^{*} \hat{A}_\mu^\dagger ] \Big\} . \label{eq:rho_0_A_decomp} \end{equation} The traceless property of each $\hat{A}_\mu$ ensures that the density matrix has unit trace, while the choice of the coefficients $f_\mu$ must be consistent with the non-negativity of the density matrix. Using Eqs. \ref{eq:Phi_M_lightcone} and \ref{eq:Phi_eigs}, the time-evolved reduced density matrix along the lightcone is: \begin{eqnarray} \hat{\rho}'_{{\rm out},M} (t) &\stackrel{(t=N)}{=}& \Phi_M^{N-M+1}[\hat{\rho}'_{{\rm in},M} (0)] \\ &=& \frac{1}{d^M} \Big\{ \hat{\mathbb{I}}_d^{\otimes M} + \sum_\mu [ z_\mu^{N-M+1} f_\mu \hat{A}_\mu + {\rm h.c.} ] \Big\} , \nonumber \label{eq:rho_n_t} \end{eqnarray} which is a convenient solution in terms of the eigenvalues $z_\mu$ and eigenoperators $\hat{A}_\mu$ of the quantum channel $\Phi_M$.


\section{Regimes of lossy and lossless information transfer}
\label{sec:classes}

\begin{figure*}[t]
    \centering
        \includegraphics[width=\linewidth]{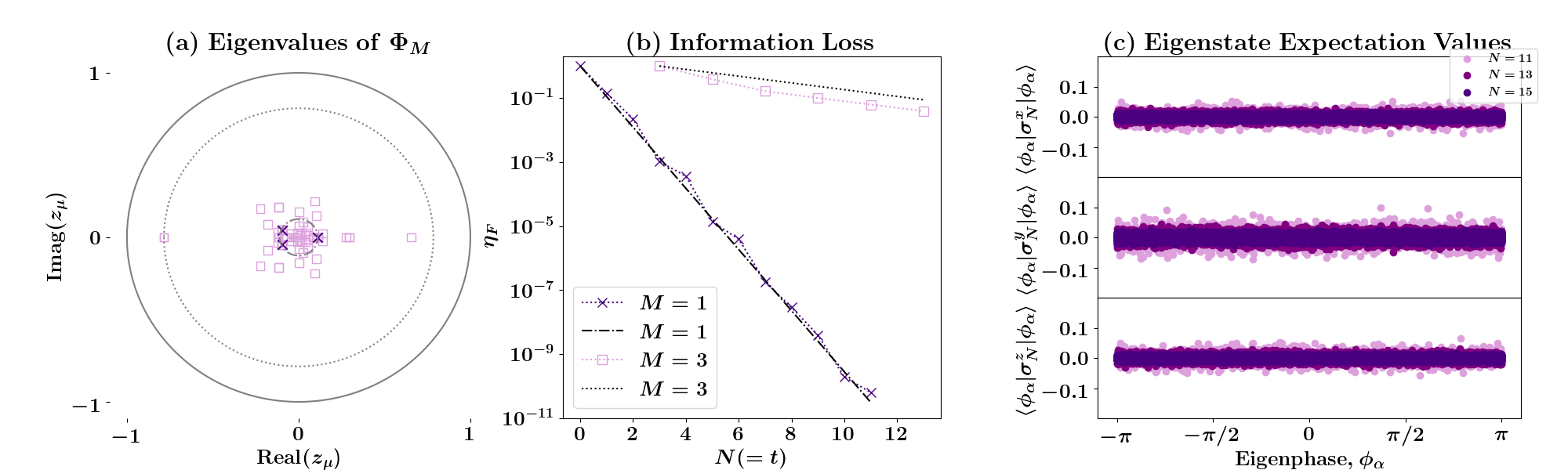}
    \caption{(a): The non-trivial eigenvalues $z_\mu$ of $\Phi_1 (\hat{U})$ and $\Phi_3 (\hat{U})$ for a Haar-random unitary $\hat{U}$. All the eigenvalues lie inside the circle, i.e., $|z_\mu |< 1$. 
    (b): Since there are no peripheral eigenvalues, the evolution along the lighcone exhibits an unavoidable loss of information. The results are for a continuous parameter $\lambda$ encoded in an input state $\hat{\rho}_{{\rm in},1}^{(\lambda)} (0) = e^{i \lambda \sigma^z}\ket{+}\bra{+}e^{-i \lambda \sigma^z}$ (for $M=1$), and $\hat{\rho}_{{\rm in},3}^{(\lambda)} (0) = \left(e^{i \lambda \sigma^z}\ket{+}\bra{+}e^{-i \lambda \sigma^z}\right)^{\otimes 3}$ (for $M=3$), where $\ket{+} = (\ket{0} + \ket{1})/\sqrt{2}$. The decay rate $\eta_F \sim |z_{\rm max}|^{t}$ is represented with the dash-dotted and dotted lines, respectively, for $M=1$ and $M=3$. 
    (c) Expectation values of the local Pauli operators at the last site of the chain, with respect to the eigenstates $\ket{\phi_\alpha}$ of the global Floquet unitary $\mathbb{U}$ [composed using the same local unitary gate $\hat{U}$ as in (a) and (b)]. }
    \label{fig:lossy_remote_sensing}
\end{figure*}

Having shown in the previous section that the $M$-qudit reduced density matrix $\hat{\rho}_{{\rm out}, M}(t)$ can be computed efficiently on the lightcone ($t=N$) and for small enough $M$, we now turn to the problem of information transfer through the chain. Central to the discussion henceforth is the following result:

\begin{result} \label{result:classes}
From the perspective of information transfer through the qudit chain, the quantum channels $\Phi_M$ may be divided into two classes:
\begin{enumerate}[label=(\roman*)]
\item \label{item:lossy} Channels for which all non-trivial eigenvalues have modulus less than one, $|z_\mu|< 1$. This leads unavoidably to loss of information along the lightcone.
\item \label{item:lossless} Channels for which there are some non-trivial eigenvalues that have modulus equal to one, $|z_\mu| = 1$ (called \emph{peripheral eigenvalues}). Lossless information transfer is possible in this case, depending on the encoding of the information in the initial state.
\end{enumerate}
\end{result} 
This result was reported in several recent works, in the context of quantum and classical channel capacities under Markovian noise channels \cite{Faw-25a,Sin-25a,Sin-25-arxiv}. Here, for completeness, we sketch a proof specific to our setting, focussing on the case of discrete information encoded in one of two initial states $\hat{\rho}^{\prime (1)}_{{\rm in},M} (0)$ or $\hat{\rho}^{\prime (2)}_{{\rm in},M} (0)$ [see Appendix \ref{app:c} for the corresponding argument for the case of continuous information encoded in a $\lambda$-parameterised state $\hat{\rho}^{\prime (\lambda)}_{{\rm in},M} (0)$].

Result \ref{result:classes}\ref{item:lossy} follows from the form of Eq. \ref{eq:rho_n_t} and the definition of the trace distance. According to the definition in Eq. \ref{eq:trace_distance}, the trace distance between the two possible time-evolved output states $\hat{\rho}^{\prime (1)}_{{\rm out},M} (N)$ and $\hat{\rho}^{\prime (2)}_{{\rm out},M} (N)$ is the trace norm of the difference: \begin{equation} \hat{\rho}^{\prime (1)}_{{\rm out},M} (N) - \hat{\rho}^{\prime (2)}_{{\rm out},M} (N) = \sum_\mu [z_\mu^{N-M+1} (f_\mu^{(1)} - f_{\mu}^{(2)})\hat{A}_\mu + {\rm h.c.}] , \end{equation} which decays at least as quickly as $|z_{\rm max}|^{N-M+1}$ where $z_{\rm max}$ is the principal eigenvalue (the non-trivial eigenvalue with the largest absolute value). The loss of information during transmission through the chain therefore also decays at least as quickly as $\eta_D \sim |z_{\rm max}|^{N-M+1}$. Another way to see this is to notice that, if $|z_\mu| < 1$ for all $\mu$, both output density matrices decay to the thermal state $\hat{\rho}^{\prime (1|2)}_{{\rm out},M} (N) \stackrel{N\to\infty}{\longrightarrow} \hat{\mathbb{I}}_d^{\otimes M} / d^M $ in the long time / long chain limit and become indistinguishable.

To see Result \ref{result:classes}\ref{item:lossless}, let us suppose that the initial state is in the form of Eq. \ref{eq:rho_0_A_decomp}, but with the sum restricted to values of the index $\mu$ for which the eigenvalue has unit modulus $z_\mu = e^{i\phi_\mu}$. Then the time-evolved density matrix along the lightcone is: 
\begin{eqnarray} \hat{\rho}'_{{\rm out},M} (N) &=& \frac{1}{d^M} \Big\{ \hat{\mathbb{I}}_d^{\otimes M} + \label{eq:peripeheral_rho_n_t} \label{eq:rho_peripheral} \\ && + \sum_{\mu, |z_\mu|=1} [ e^{i\phi_\mu (N-M+1)} f_\mu \hat{A}_\mu + {\rm h.c.} ] \Big\} .  \nonumber \end{eqnarray}
If the phases $\phi_\mu$ in Eq. \ref{eq:rho_peripheral} are all equal to zero, then the output $M$-qudit state is identical to the input $M$-qudit state $\hat{\rho}'_{{\rm out},M} (N) = \hat{\rho}'_{{\rm in},M}(0)$ and all information is perfectly preserved as it passes through the chain. If there are some non-zero phases $\phi_\mu \neq 0$, the output reduced density matrix is not stationary as a function of $t = N$. However, due to the Poincar\'{e} recurrence theorem, there must be times $t = N = N_*$ for which any output density matrix in Eq. \ref{eq:peripeheral_rho_n_t} returns arbitrarily close to its initial state $\hat{\rho}'_{{\rm out},M} (N_*) = \Phi_M^{N_* - M + 1}[\hat{\rho}'_{{\rm in},M}(0)] \approx \hat{\rho}'_{{\rm in},M}(0)$. At such times we have:
\begin{equation} D[\hat{\rho}^{\prime (1)}_{{\rm out},M}(N_*), \hat{\rho}^{\prime (2)}_{{\rm out},M}(N_*)] \approx D[\hat{\rho}^{\prime (1)}_{{\rm in},M}(0), \hat{\rho}^{\prime (2)}_{{\rm in},M}(0)] , \end{equation} and so $\eta_D (N_*, N_*, M) \approx 1$. However, the trace distance cannot increase under the quantum channel $\Phi_M$. So we must have 
\begin{equation} D[\hat{\rho}^{\prime (1)}_{{\rm out},M}(N), \hat{\rho}^{\prime (2)}_{{\rm out},M}(N)] \approx D[\hat{\rho}^{\prime (1)}_{{\rm in},M}(0), \hat{\rho}^{\prime (2)}_{{\rm in},M}(0)] , \end{equation} and $\eta_D (N, N, M) \approx 1$ for all possible spacetime points along the lightcone $t = N$. This shows that the information encoded in the eigenoperators associated with peripheral eigenvalues is preserved along the light cone of the brickwork circuit.

The simplest example of lossless remote sensing is when the two-qudit gate $\hat{U}$ is the \texttt{SWAP} operator, $\hat{S}$, which operator acts as $\hat{S} (\hat{\mathcal{O}} \otimes \hat{\mathcal{O}}') \hat{S} = \hat{\mathcal{O}}' \otimes \hat{\mathcal{O}}$ on arbitrary single-qudit operators $\hat{\mathcal{O}}$ and $\hat{\mathcal{O}}'$. Using this property, it is easy to see that for for any $M$-qudit input operator $\hat{A}$, we have $\Phi_M [\hat{A}] = \hat{A}$, i.e., all eigenvalues of the quantum channel are $z_\mu = 1$. By Result \ref{result:classes}\ref{item:lossless} we therefore expect lossless information transfer. Indeed, referring to Fig. \ref{fig:schematic}, it is easy to see that the action of the \texttt{SWAP} brickwork circuit is to permute the initial reduced density matrix of the first $M$ qudits through the chain to the last $M$ sites after an evolution time $t = N$, i.e., 
all information is perfectly transmitted through the circuit without any loss, $\eta_D = 1$. The \texttt{SWAP} circuit is non-interacting and integrable, so lossless remote sensing is not particularly surprising for this example. For a \emph{generic} two-qudit gate $\hat{U}$, information typically degrades as it propagates through the chain, as we show in the next section.



\section{Information loss through generic brickwork circuits} \label{sec:generic_loss}

\begin{figure}[b]
    \centering
    \includegraphics[width=\linewidth]{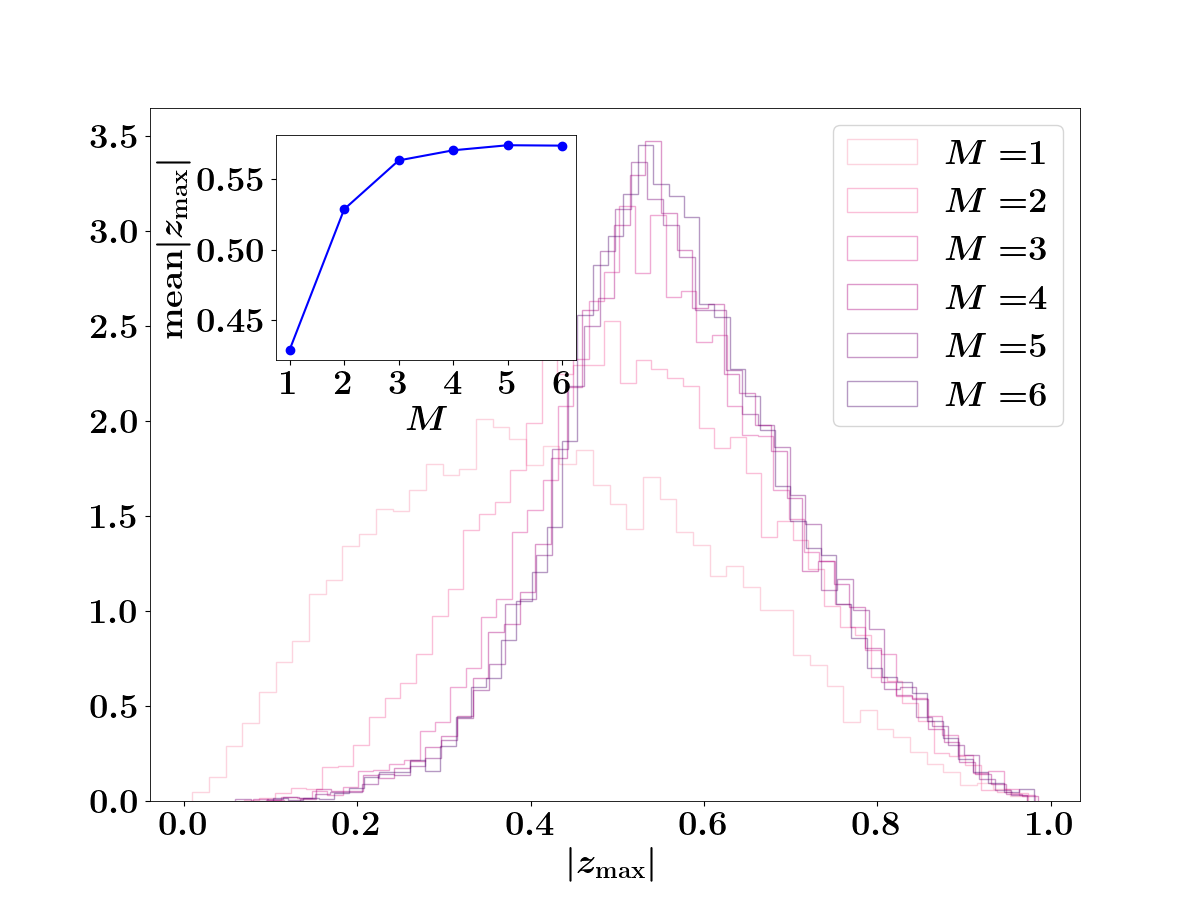}
    \caption{Distribution of $|z_{\rm max}|$ (the non-trivial eigenvalue of $\Phi_M (\hat{U})$ with largest magnitude) for an ensemble of $10000$ Haar-random unitaries $\hat{U}$. Peripheral eigenvalues ($|z_{\rm max}|=1$) are extremely rare. The inset shows the mean of each distribution as a function of $M$, which saturates at a value smaller than unity as $M$ increases.}
    \label{fig:z_distribution}
\end{figure}

Result \ref{result:classes} shows that the presence or absence of peripheral eigenvalues in the spectrum $\Phi_M$ is the property that determines whether lossless information transfer on the lightcone is possible or not. In this section we show that, for a generic two-qudit unitary $\hat{U}$, the corresponding quantum channel $\Phi_M (\hat{U})$ typically does not have peripheral eigenvalues, so that information degrades as it propagates through the chain.

To illustrate this, we choose a two-qudit unitary $\hat{U}$ at random from the circular unitary ensemble (CUE) and we compute the spectrum $\{ z_\mu \}$ of the corresponding quantum channel $\Phi_M (\hat{U})$, shown in Fig.~\ref{fig:lossy_remote_sensing}(a). We see that for $M \in \{ 1,3 \}$ the magnitude of the principal eigenvalue is smaller than unity $|z_{\rm max}| < 1$. By Result \ref{result:classes}\ref{item:lossy} we therefore expect unavoidable loss of information during the dynamics through the chain. This is confirmed in Fig.~\ref{fig:lossy_remote_sensing}(b) where, for a continuous parameter $\lambda$ encoded in an initial state $\hat{\rho}_{{\rm in},M}^{(\lambda)} (0)$, we compute the loss $\eta_F$ of the QFI due to evolution through the chain. As expected, $\eta_F$ decays exponentially as a function of $t = N$, at the rate $\eta_F \sim |z_{\rm max}|^t$.

This lack of peripheral eigenvalues is not specific to the particular instance of the random two-qudit unitary $\hat{U}$ chosen in this example. In Fig.~\ref{fig:z_distribution} we plot the distribution of $|z_{{\rm max}}|$ for the quantum channel $\Phi_M (\hat{U})$, corresponding to many random instances of the two-qudit unitary $\hat{U}$. For $M \in \{ 1,2,...,6 \}$ we see that peripheral eigenvalues $|z_{\rm max}| = 1$ are extremely rare. Also, as $M$ increases, the mean $\overline{|z_{\rm max}|}$ of the distribution of principal eigenvalues does not approach unity, but saturates at a value well below unity (see inset to Fig.~\ref{fig:z_distribution}). A similar phenomenon has been observed for a related class of quantum channels and is connected with Reulle-Pollicot resonances \cite{Pol-85a, Rue-86a, Mor-24a, Zni-24a}.

\begin{figure}
    \centering
    \includegraphics[width=0.8\linewidth]{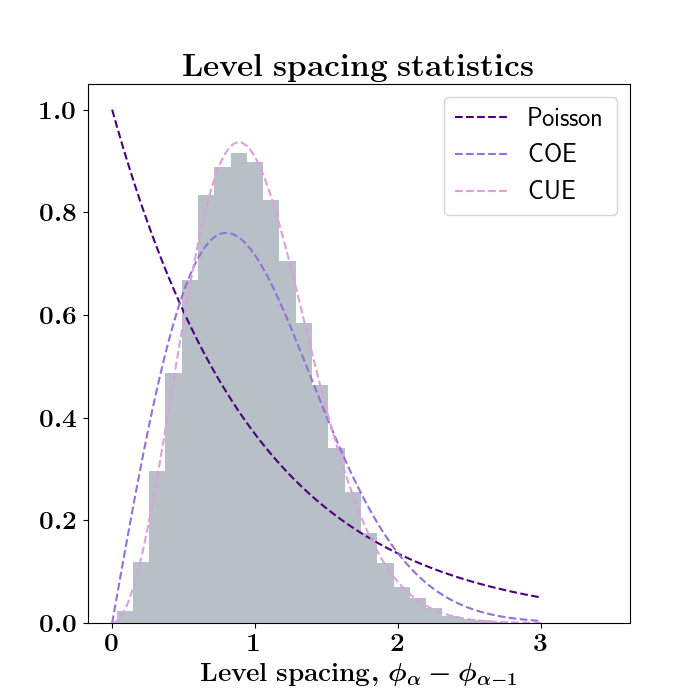}
    \caption{Level spacing statistics of the Floquet unitary $\hat{\mathbb{U}}$, built with a Haar-random local unitary gate $\hat{U}$. The distribution of level spacings is consistent with that of the circular unitary ensemble (CUE). }
    \label{fig:haar_levelspacing}
\end{figure}

Finally, we note that for a random two-qudit gate $\hat{U}$ the corresponding global Floquet unitary $\hat{\mathbb{U}}$ is typically non-integrable and thermalising. The non-integrability is confirmed by the level spacing statistics of the Floquet eigenphases $\{ \phi_\alpha \}$, defined via $\hat{\mathbb{U}} \ket{\phi_\alpha} = e^{i\phi_\alpha} \ket{\phi_\alpha}$, which are shown in Fig. \ref{fig:haar_levelspacing}. The distribution follows that of the CUE for matrices of dimension $d^{N+1} \times d^{N+1}$, consistent with quantum chaos. Moreover, in Fig. \ref{fig:lossy_remote_sensing}(c) we plot the eigenstate expectation values $\langle \phi_\alpha | \hat{\sigma}_N^\mu | \phi_\alpha \rangle$ for Pauli operators on the final qubit (assuming $d=2$). These values cluster tightly around zero (the thermal expectation) and the spread around zero decreases with increasing system size $N$. This indicates that the Floquet eigenstates satisfy the eigenstate thermalisation hypothesis (ETH) with respect to local observables at the chain’s end. As a result, the long-time averages of such observables converge to their thermal values \cite{Rig-08,DAl-16,Mor-18}, implying that the final spin effectively thermalises.

In light of the model being chaotic and locally thermalising, it is not surprising that information is lost during transmission through the chain. However, later in Sec. \ref{sec:example} we will see a more surprising example of lossless information transfer through a nonintegrable circuit.



\section{Lossless information transfer}
\label{sec:lossless_info_transfer}


It is clear from Result \ref{result:classes} that lossless remote sensing along the lightcone is only possible if the quantum channel $\Phi_M$ has peripheral eigenvalues. To explore this further, in this section we focus on some of the simplest quantum channels $\Phi_M$ for $M=1$ or $M=2$, acting on a system of qubits ($d=2$) or qutrits ($d=3$). The section is structured as follows.

First, in Sec. \ref{sec:d2_M1}, we consider the $M=1$ channel $\Phi_1 (\hat{U})$ acting on a qubit ($d=2$). We show that $\Phi_1 (\hat{U})$ has peripheral eigenvalues only if the two-qubit unitary $\hat{U}$ is dual-unitary, which means that it is not only unitary as a propagator in the time direction, but is also unitary when interpreted as a propagator in the spatial direction. 

Next, in Sec. \ref{sec:d2_M2} we consider the $M=2$ channel $\Phi_2 (\hat{U})$, again acting on a system of qubits $d=2$. We show numerically that $\Phi_2(\hat{U})$ can have peripheral eigenvalues even if $\hat{U}$ is not dual-unitary. We conjecture a necessary condition which must be obeyed by $\hat{U}$ for $\Phi_2 (\hat{U})$ to exhibit peripheral eigenvalues. 

Finally, in Sec. \ref{sec:d3_M1} we consider the $M=1$ channel $\Phi_1$ acting on a qutrit $d=3$. In this case too, we show numerically that the channel can exhibit peripheral eigenvalues even if the two-qutrit unitary $\hat{U}$ is not dual-unitary.

\subsection{Lossless information transfer for $d=2$ and $M=1$} \label{sec:d2_M1}

In the Pauli operator basis, the non-trivial part of the quantum channel $\Phi_1$ defined in Eq. \ref{eq:Phi_1} can be represented as a $3 \times 3$ real matrix with the elements: \begin{equation} \Phi_1^{\mu\nu}(\hat{U}) = \frac{1}{4} {\rm Tr} [(\hat{\mathbb{I}}_2 \otimes \hat{\sigma}^\mu) \hat{U} (\hat{\sigma}^\nu \otimes \hat{\mathbb{I}}_2 ) \hat{U}^\dagger] , \label{eq:Phi_matrix_elements_1} \end{equation} where $\{ \hat{\sigma}^\mu \}_{\mu \in \{x,y,z \}}$ are the qubit Pauli operators. The qubit channel $\Phi_1$ is completely determined by the choice of unitary $\hat{U}$ which, for two qubits, can always be expressed in the form \cite{Kha-01a,She-04a}:
\begin{equation} \hat{U} = (\hat{u}' \otimes \hat{u}) \hat{V}(J^x, J^y, J^z) (\hat{v} \otimes \hat{v}') , \label{eq:U_two_qubit} \end{equation} where $\hat{u}$, $\hat{u}'$, $\hat{v}$ and $\hat{v}'$ are single-qubit unitaries, and:
\begin{equation}\label{eq:two_qubit_nonlocal_param} \hat{V}(J^x, J^y, J^z) = e^{iJ^x \hat{\sigma}^x \otimes \hat{\sigma}^x + iJ^y \hat{\sigma}^y \otimes \hat{\sigma}^y + iJ^z \hat{\sigma}^z \otimes \hat{\sigma}^z } , \end{equation} where we assume $-\pi/2 \leq J^\mu \leq \pi/2$. It is also known that by fixing any two of the parameters $\{ J^x, J^y, J^z \}$ to the value $|J^\mu| = \pi/4$ the two-qubit operator $\hat{U}$ is dual-unitary \cite{Ber-19a}, which means that $\hat{U}$ is not only unitary as a propagator in the time direction, but is also unitary when interpreted as a propagator in the spatial direction. 

The following result was first proved in Ref. \cite{Ber-20c}. However, here we provide an alternative, arguably more elementary, proof.

\begin{result}\label{result:qubit_channel} 
The qubit channel $\Phi_1(\hat{U})$ has peripheral eigenvalues only if the two-qubit operator $\hat{U}$ is dual-unitary.
\end{result}

To prove Result \ref{result:qubit_channel}, we first substitute the expression for $\hat{U}$ in Eq. \ref{eq:U_two_qubit} into Eq. \ref{eq:Phi_matrix_elements_1}, to obtain the channel matrix elements: 
 \begin{equation} \Phi_1^{\mu\nu}(\hat{U}) = \frac{1}{d^2} {\rm Tr} [(\hat{\mathbb{I}}_2 \otimes \hat{u}^\dagger\hat{\sigma}^\mu \hat{u}) \hat{V} (\hat{v}\hat{\sigma}^\nu \hat{v}^\dagger \otimes \hat{\mathbb{I}}_2 ) \hat{V}^\dagger] . \label{eq:Phi_matrix_elements_2} \end{equation} We can rewrite the unitarily transformed Pauli operators as $\hat{u}^\dagger \hat{\sigma}^\mu \hat{u} = \sum_{\xi} Q_{\mu\xi} \hat{\sigma}^\xi$ and $\hat{v} \hat{\sigma}^\nu \hat{v}^\dagger = \sum_{\xi} R_{\nu\xi} \hat{\sigma}^{\xi}$, where $Q, R \in SO(3)$. Substituting these into Eq. \ref{eq:Phi_matrix_elements_2} gives: 
\begin{equation} \Phi_1^{\mu\nu}(\hat{U}) = \sum_{\xi\xi'} Q_{\mu\xi} \Phi_{\xi\xi'}(\hat{V}) R^T_{\xi'\nu} , \label{eq:Phi_matrix_elements_SVD} \end{equation} which is now expressed in terms of the matrix elements $\Phi_1^{\xi\xi'}(\hat{V})$ of the quantum channel with respect to the two-qubit unitary $\hat{V}$. It is a straightforward calculation (see Appendix \ref{app:d}) to show that $\Phi_1^{\xi\xi'}(\hat{V})$ is a $3 \times 3$ diagonal matrix, with the diagonal elements: 
\begin{eqnarray} \Phi_1^{xx}(\hat{V}) &=& \sin(2J^y)\sin(2J^z)  , \label{eq:Phi_1^XX} \\ \Phi_1^{yy}(\hat{V}) &=& \sin(2J^z)\sin(2J^x)  , \\ \Phi_1^{zz}(\hat{V}) &=& \sin(2J^x)\sin(2J^y)  . \label{eq:Phi_1^ZZ} \end{eqnarray} In other words, Eq. \ref{eq:Phi_matrix_elements_SVD} is essentially a singular value decomposition of $\Phi_1 (\hat{U})$, with the singular values $|\Phi_1^{\xi\xi}(\hat{V})|$. 
Now, with the singular values of $\Phi_1(\hat{U})$ in hand, we can apply a result known as Weyl's inequality, which says that $s_1 s_2 \hdots s_k \geq |z_1 z_2 \hdots z_k|$ where $s_1 \geq s_2 \geq s_{3}$ are the singular values of $\Phi_1(\hat{U})$ in decreasing order, $|z_1| \geq |z_2| \geq |z_3|$ are its eigenvalues' magnitudes, and $k \in \{1,2,3 \}$. It follows that the eigenvalue of $\Phi_1(\hat{U})$ with the largest modulus must be less than or equal to the largest singular value, $s_1 \geq |z_1|$. If at least two of $\{ J^x, J^y, J^z\}$ satisfy $|J^\mu| \neq \pi/4$ then we see that all singular values in Eq. \ref{eq:Phi_1^XX}-\ref{eq:Phi_1^ZZ} must be less than one and therefore all eigenvalues must have modulus smaller than one. Conversely, at least two of $\{ J^x, J^y, J^z\}$ must obey $|J^\mu| = \pi/4$ (i.e., $\hat{U}$ must be dual-unitary) to have a non-trivial eigenvalue with unit modulus.

It now follows from Result \ref{result:classes} and Result \ref{result:qubit_channel} that:
\begin{corollary} \label{cor:lossless_qubit_chain}
For a chain of qubits evolving by a brickwork circuit, lossless information transfer along the $M=1$ lightcone is only possible for dual-unitary circuits.
\end{corollary}

\subsection{Lossless information transfer for $d=2$ and $M=2$} \label{sec:d2_M2}

It is natural to ask whether the necessary condition for lossless information transfer along the $M=1$ lightcone, given in Corollary \ref{cor:lossless_qubit_chain}, applies also to $M > 1$. In this section we examine this numerically, focussing on the case of $M=2$, again for a chain of qubits ($d=2$).

Specifically, the evolution along the lightcone of the initial state of the first $M=2$ qubits is given by repeated application of the $M=2$ quantum channel, as shown in Eq. \ref{eq:rho_N-1,N}. A $16 \times 16$ matrix representation of the channel $\Phi_2$ is given by:
\begin{equation}
    \Phi_2^{\mu\nu} = \frac{1}{2} \Tr \{ (\hat{\mathbb{I}}_2 \otimes \hat{\Gamma}^{\mu}) \mathcal{U}_3 [\hat{\Gamma}^{\nu} \otimes \hat{\mathbb{I}}_2] \} ,
\end{equation}
where $\mathcal{U}_3$ is the three-qubit unitary channel defined in Eq. \ref{eq:U_3} and $\{ \hat{\Gamma}^\mu \}_{\mu=0}^{15}$ is a Hermitian operator basis for two qubits that is orthonormal ${\rm Tr}[\hat{\Gamma}^\mu \hat{\Gamma}^\nu] = \delta_{\mu\nu}$.


\begin{figure}[t]
  \centering
  \includegraphics[width=\columnwidth]{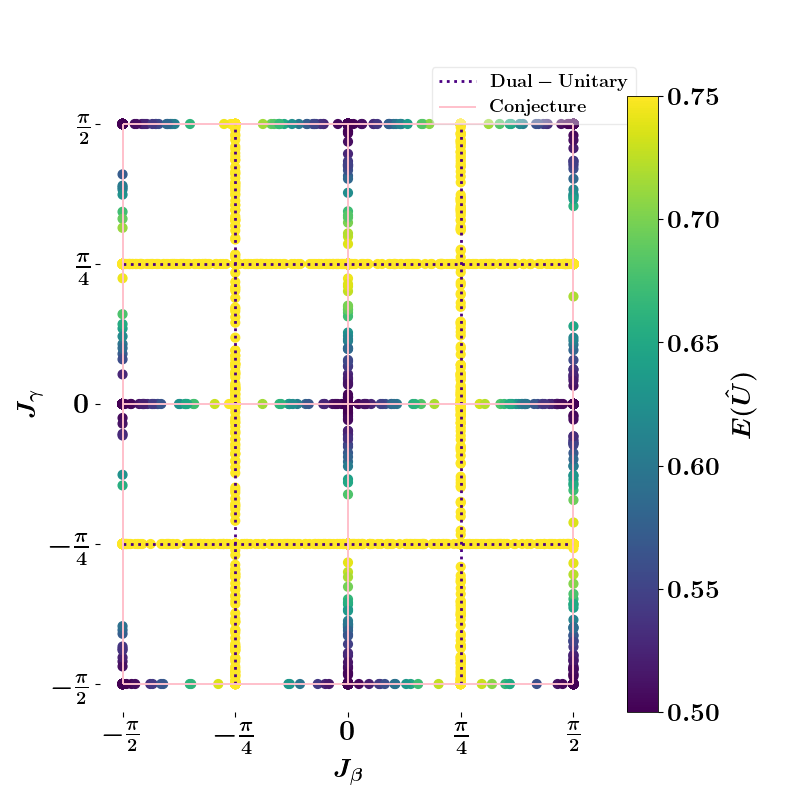}
  \caption{ Numerical optimization of the nonlocal parameters, characterizing the two-qubit unitary, for the peripheral eigenvalues of the map $\Phi_2$ composed of two qubit unitary. The conjecture states that $J_\alpha= \frac{\pi}{4} + n\pi$, and the other two $J_{\beta}$, $J_{\gamma}$ are on the x-axis and y-axis, respectively, for $\alpha\neq\beta\neq\gamma$. 
  The dotted pink lines represent at least one of $J_\alpha$ and $J_\beta$ assuming a value of $\frac{\pi}{4} + n\pi$, i.e., a dual-unitary parameter. The solid line represents a unitary achieving peripheral eigenvalues for $\Phi_2$. All the points on the dual-unitary line have maximum operator entanglement for $d=2$, on the other hand, the conjecture has less operator entanglement.}
  \label{fig:twolocalinput_result}
\end{figure}
\noindent

Following Result \ref{result:classes}, the presence or absence of peripheral eigenvalues of the map $\Phi_2$ determine whether lossless remote sensing is achievable along the lightcone. Since the spectrum of any quantum channel $\Phi_M (\hat{U})$ is a subset of the spectrum of the channel $\Phi_{M+1} (\hat{U})$, we know that all two-qubit dual-unitary operators $\hat{U}$ that result in peripheral eigenvalues of $\Phi_1 (\hat{U})$ will also give peripheral eigenvalues of $\Phi_2 (\hat{U})$. We would like to know if there are additional, \emph{non-dual-unitary} $\hat{U}$ that give peripheral eigenvalues for $\Phi_2 (\hat{U})$ [but not for $\Phi_1 (\hat{U})$].

Although the extension from $\Phi_1 (\hat{U})$ to $\Phi_2 (\hat{U})$ represents only a relatively minor increase in formal complexity, an analytic characterization of a sufficient condition for peripheral eigenvalues of $\Phi_2 (\hat{U{}})$ (similar to Result \ref{result:qubit_channel}) appears to be a much more difficult problem. We therefore adopt a numerical approach, optimising the parameters $\{J^x, J^y, J^z\}$ and the local unitaries $u$, $u'$, $v$, and $v'$ in the parameterisation of $\hat{U}$ given in Eq. \ref{eq:U_two_qubit}, in order to identify instances when $\Phi_2 (\hat{U})$ exhibits peripheral eigenvalues.

The numerical results, as expected, reveal instances of dual-unitary operators $\hat{U}$ that achieve the peripheral eigenvalue for $\Phi_2 (\hat{U})$ (i.e., two of the parameters $\{ J^x, J^y, J^z \}$ satisfy $|J^\mu| = \pi/4$). However, we also identify instances when $\Phi_2 (\hat{U})$ has peripheral eigenvalues but $\hat{U}$ is not dual-unitary. In all optimised instances exhibiting peripheral eigenvalues, we find that at least \emph{one} of the parameters $\{ J^x, J^y, J^z \}$ satisfies $|J^\mu| = \pi/4$. This suggests a convenient visualization of our results: we ignore the value that obeys $|J^\mu| = \pi/4$ and plot the remaining two values $\{ J^\beta, J^\gamma \}$ on the plane. The results are presented in Fig. \ref{fig:twolocalinput_result}. The dual-unitaries are the points for which at least one of the remaining $\{ J^\beta, J^\gamma \}$ obeys $|J^{\beta | \gamma}| = \pi/4$. Fig. \ref{fig:twolocalinput_result} shows that there are many unitaries $\hat{U}$ that are not dual-unitary but lead to peripheral eigenvalues of $\Phi_2 (\hat{U})$. However, even these optimised $\{ J^\beta, J^\gamma \}$ values appear to follow a pattern, which leads us to the following conjecture:

\begin{conjecture} \label{conj:M2_d2}
The two-qubit quantum channel $\Phi_2 (\hat{U})$ has peripheral eigenvalues only if one of the parameters $\{ J^x, J^y, J^z \}$ (of the two-qubit unitary $\hat{U}$)
obeys $|J^\mu| = \pi/4$ and another obeys $|J^\beta| = k \pi/4$ for $k \in \mathbb{Z}$.
\end{conjecture}
This clearly includes the dual-unitary instances when $k=1$, but also non dual-unitary instances when $k = 0, 2$ and only one of $\{ J^x ,J^y, J^z \}$ obeys $|J^\mu| = \pi/4$. 

Another way of characterising dual-unitarity (which is valid not only for qubits but also for $d>2$) is to compute the operator entanglement of the two-qubit gate $\hat{U}$, which is maximal if and only if $\hat{U}$ is dual-unitary \cite{Zyc-04a, Ber-20b,Rat-20a}. We quantify the operator entanglement with the operator linear entropy: 
\begin{equation} E(\hat{U}) = 1 - \sum_{\mu} a_\mu^2 , \end{equation} where $a_\mu$ are the coefficients in the Schmidt decomposition $\hat{U} = \sum_\mu \sqrt{a_\mu} \hat{U}_\mu^A \otimes \hat{U}_\mu^B$. The two-qubit unitary $\hat{U}$ is dual-unitary if and only if the operator linear entropy achieves maximal its maximum value $E(\hat{U}_{\rm dual-unitary}) = 3/4$ [or, more generally, $E(\hat{U}_{\rm dual-unitary}) = \frac{d^2 - 1}{d^2}$ for a two-qudit unitary $\hat{U}_{\rm dual-unitary}$], which gives the following inequality: 
\begin{equation} 0 \leq E(\hat{U}) \leq E(\hat{U}_{\rm dual-unitary}) . \end{equation}
In Fig. \ref{fig:twolocalinput_result}, the colours of the data points represent the operator linear entropy of the optimised two-qubit unitary $\hat{U}$. The dual-unitary examples are clearly visible in yellow, but also the non dual-unitary examples, with sub-maximal values $E(\hat{U}) < 3/4$ of the operator linear entropy. Interestingly, our numerical search has not found any examples of $\Phi_2 (\hat{U})$ with peripheral eigenvalues, for which the operator linear entropy of $\hat{U}$ has a value less than $0.5$, which suggests that a minimum (but non-maximal) operator entanglement is required to give peripheral eigenvalues of $\Phi_2(\hat{U})$ (i.e., to facilitate lossless information transfer through the chain).



 

\subsection{Lossless information transfer for $d=3$ and $M=1$} \label{sec:d3_M1}

In Sec. \ref{sec:d2_M1} we proved that the two-qubit unitary $\hat{U}$ must be dual-unitary for $\Phi_1 (\hat{U})$ to exhibit peripheral eigenvalues (and therefore lossless information transfer along the lightcone). In Sec. \ref{sec:d2_M2} we showed that dual-unitarity is not necessary for the two-qubit channel $\Phi_2 (\hat{U})$ to have peripheral eigenvalues. In this section, we demonstrate that the restriction to dual-unitary $\hat{U}$ is also not necessary for the channel $\Phi_1 (\hat{U})$ acting on a qutrit $d=3$.

A $9 \times 9$ matrix representation for the qutrit quantum channel $\Phi_1$ is given by:
\begin{equation} \Phi^{\mu\nu}_1 (\hat{U}) = \frac{1}{3}
{\rm Tr} [(\hat{\mathbb{I}}_3 \otimes \hat{\Gamma}^{\mu}) \hat{U} (\hat{\Gamma}^\nu \otimes \hat{\mathbb{I}}_3 ) \hat{U}^\dagger] , \label{eq:Phi_matrix_elements_qutrit} \end{equation}
where $\{ \Gamma^\mu \}_{\mu=0}^8$ is an operator basis for a qutrit, which we assume is Hermitian $\hat{\Gamma}^{\mu \dagger} = \hat{\Gamma}^{\mu}$ and orthonormal ${\rm Tr}[\hat{\Gamma}^\mu \hat{\Gamma}^\nu] = \delta_{\mu\nu}$.

For simplicity, we consider a two-qutrit unitary of the form \begin{equation} \hat{U} =( \hat{u}' \otimes \hat{u}) \exp\left(-i\sum_{\mu=1}^8 J^\mu \hat{\Gamma}^\mu \otimes \hat{\Gamma}^\mu\right) ( \hat{v} \otimes \hat{v}') , \end{equation} where $\hat{u}, \hat{u}', \hat{v}, \hat{v}'$ are single-qutrit unitaries. This particular form does not cover the entire space of two-qutrit unitaries but allows us to work with a reduced number of parameters, which is convenient for numerical optimisation. Moreover, the cyclicity of trace ensures the quantum channel $\Phi_1^{\mu\nu}$ in Eq. \ref{eq:Phi_matrix_elements_qutrit} is independent of $\hat{u}'$ and $\hat{v}'$, and depends on a total of $24$ parameters: $8$ values of $\{ J^\mu \}_{\mu=1}^8$ in the entangling part of the unitary, and $8$ parameters each for the single-qutrit unitaries $\hat{u}$ and $\hat{v}$. We perform a numerical optimisation over these parameters, in order to identify instances when $\Phi_1 (\hat{U})$ exhibits peripheral eigenvalues. 

We note that there is no known complete parametrization of two-qutrit dual-unitaries (unlike two-qubit dual-unitaries, for which the complete parametisation is given by Eq. \ref{eq:U_two_qubit} with two of $\{J^x, J^y, J^z \}$ set to $|J^\mu| = \pi/4$). However, we can check the dual-unitarity of $\hat{U}$ by computing its operator linear entropy which attains its maximal value $E(\hat{U}) = 8/9$ if and only if $\hat{U}$ is dual-unitary.

\begin{figure}[t]
  \centering
  \includegraphics[width=0.9\columnwidth]{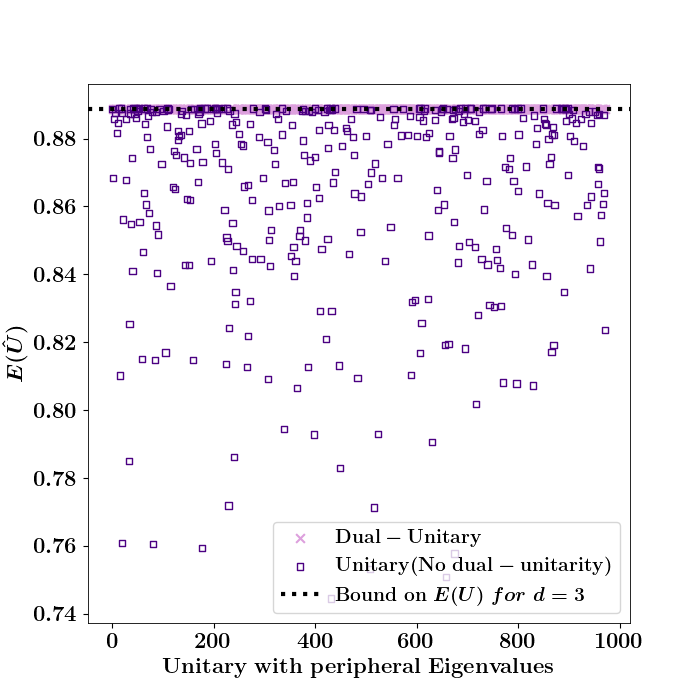}
  \caption{Optimized two-qutrit unitary with peripheral eigenvalues for $\Phi_1$. The optimization routine returns several unitary operators that do not appear to be dual-unitary. This can be verified by the linear entropy of the unitaries, many unitaries do not approach the maximum linear entropy, a necessary characteristic of dual-unitary. The optimization routine also returned dual-unitary operators (in pink), achieving peripheral eigenvalues.
  }
  \label{fig:qutrit_distribution}
\end{figure}

In Fig. \ref{fig:qutrit_distribution} we plot the operator linear entropy $E(\hat{U})$ corresponding to a number of optimised two-qutrit unitaries $\hat{U}$ which lead to peripheral eigenvalues of $\Phi_1 (\hat{U})$. Although many of the optimised examples are dual-unitary (corresponding to maximal operator linear entropy -- the horizontal dotted line), there are also many \emph{non-dual-unitary} $\hat{U}$ (operator linear entropy less than the maximum value) which nevertheless lead to peripheral eigenvalues of $\Phi_1 (\hat{U})$, and hence are capable of lossless information transfer through the qutrit chain. There are no examples in Fig. \ref{fig:qutrit_distribution} with very low operator linear entropy, which again suggests that some minimum amount is required to facilitate peripheral eigenvalues and lossless information transfer.

The analysis of this section and the previous section (\ref{sec:d2_M2}) suggests that the necessary condition of Result \ref{result:qubit_channel} should therefore be regarded as a special case in the qubit ($d=1$) and $M=1$ setting. For higher-dimensional local Hilbert spaces $d \geq 3$, or for extended operators along the lightcone $M \geq 2$, the presence of peripheral eigenvalues does not necessarily imply dual-unitarity.

\section{Example - Lossless information transfer through a chaotic and thermalising brickwork circuit}
\label{sec:example}

In Sec. \ref{sec:generic_loss}, we saw that information is typically lost when it is transmitted along the lightcone of a generic brickwork circuit. This is not surprising, since such circuits are non-integrable and satisfy the ETH, which implies thermalisation and the loss of information about the initial state. In this section, however, we provide an example that demonstrates that neither quantum chaos nor ETH necessarily precludes the possibility of lossless information transfer along the lightcone.


\begin{figure*}[t]
    \centering
    \includegraphics[width=\linewidth]{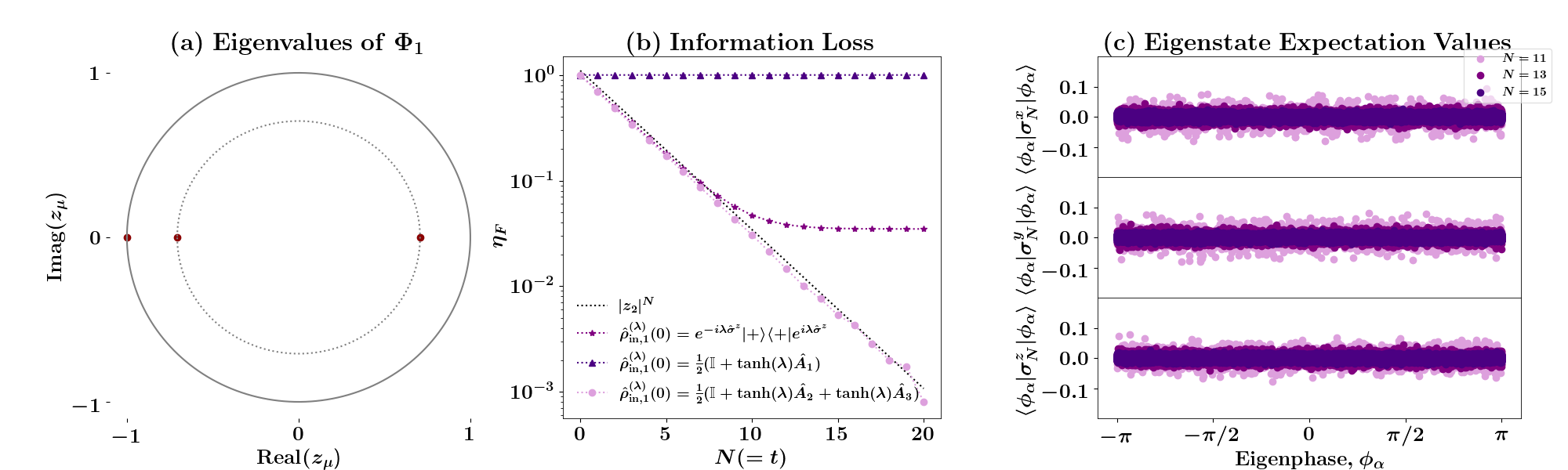}
    \caption{(a): The non-trivial eigenvalues of $\Phi_1$ as described in Eq.~\ref{eq:phi_1_erg_nonmix}, for $J_z =\pi/8$. All the eigenvalues are real, and one peripheral eigenvalue is $-1$. 
    (b): QFI is transmitted through the chain without loss for an initial state  $\rho_{\rm in, 1}^{(\lambda)}(0) = \frac{1}{2}\left[ \hat{\mathbb{I}}_2 + \tanh{\lambda} \hat{A}_1 \right]$, where $\hat{A}_1$ is the eigenoperator with peripheral eigenvalue $-1$. QFI decays exponentially, at the rate $\eta_F \sim |\sin (2J^z)|^t$ for the initial state given in Eq. \ref{eq:eta_F_loss}. For initial state $\rho_{\rm in, 1}^{(\lambda)}(0) = e^{-i\lambda \sigma^z}\ket{+}\bra{+}e^{i\lambda \sigma^z}$, the QFI decays exponentially and then saturates to a non-zero value, since the initial state has some overlap with the eigenoperator $\hat{A}_1$ associated with the peripheral eigenvalue. 
    (c) Expectation of the local Pauli operators at the last site of the chain with the eigenstates of the Floquet operator $\mathbb{U}$, composed of the same unitary $U$ as in (a) and (b).
    }
    \label{fig:lossless_remote_sensing}
\end{figure*}

We focus on the case of $M=1$ and $d=2$ (i.e., a chain of qubits). Our objective is to construct a two-qubit unitary $\hat{U}$ such that the associated quantum channel $\Phi_1(\hat{U})$ possesses exactly one non-trivial peripheral eigenvalue of unit modulus, while the remaining two lie strictly within the unit circle. According to Result~\ref{result:classes}, this structure allows lossless information transfer along the lightcone. Moreover, Result~\ref{result:qubit_channel} implies that any such $\hat{U}$ must be dual-unitary. We adopt a construction inspired by Claeys and Lamacraft~\cite{Cla-21a} to realise this scenario.

Specifically, we fix the parameters of the two-qubit unitary $\hat{U}$ in Eq.~\ref{eq:U_two_qubit} as $J^x = -J^y = -\pi/4$, ensuring dual-unitarity of $\hat{U}$, and we choose the single-qubit unitaries to have the form:
\begin{equation} \hat{u} = \hat{w} e^{-i\theta_Q \hat{\sigma}^z /2}, \qquad \hat{v} = e^{i \theta_R \hat{\sigma}^z /2} \hat{w}^\dagger, \label{eq:u_v_choice} \end{equation}
where \(\hat{w}\) is an arbitrary unitary on a single qubit. The remaining single-qubit unitaries $\hat{u}'$ and $\hat{v}'$ in Eq. \ref{eq:U_two_qubit} can be freely chosen, so we sample them at random from the CUE.

To see why this gives a quantum channel $\Phi_1 (\hat{U})$ with the desired spectral properties, we observe that substituting $J^x = -J^y = -\pi/4$ into Eq. \ref{eq:Phi_matrix_elements_SVD} gives the decomposition: \begin{equation} \Phi_1 (\hat{U}) = Q \left( \begin{array}{ccc} \sin(2J^z) & 0 & 0 \\ 0 & -\sin(2J^z) & 0 \\ 0 & 0 & -1 \end{array} \right) R^T , \end{equation} where $Q, R \in SO(3)$ are defined by the relations $\hat{u}^\dagger \hat{\sigma}^\mu \hat{u} = \sum_{\xi} Q_{\mu\xi} \hat{\sigma}^\xi$ and $\hat{v} \hat{\sigma}^\nu \hat{v}^\dagger = \sum_{\xi} R_{\nu\xi} \hat{\sigma}^{\xi}$ (see Eq. \ref{eq:Phi_matrix_elements_SVD}). If $Q$ and $R$ can be constructed to preserve the $-1$ entry of the diagonal matrix as an eigenvalue of $\Phi_1 (\hat{U})$ then we will have achieved our goal. The choice of single-qubit unitaries $\hat{u}$ and $\hat{v}$ in Eq. \ref{eq:u_v_choice} guarantee precisely this, by ensuring that $Q$ and $R$ have the form:
\begin{eqnarray} Q &=& W \left( \begin{array}{ccc} \cos\theta_Q & -\sin\theta_Q & 0 \\ \sin\theta_Q & \cos\theta_Q & 0 \\ 0 & 0 & 1 \end{array} \right) , \label{eq:Q} \\ R &=& W \left( \begin{array}{ccc} \cos\theta_R & -\sin\theta_R & 0 \\ \sin\theta_R & \cos\theta_R & 0 \\ 0 & 0 & 1 \end{array} \right) , \label{eq:R} \end{eqnarray} where $W \in SO(3)$ is an arbitrary $3 \times 3$ special orthogonal matrix. Using Eq. \ref{eq:Phi_matrix_elements_SVD}, this gives the channel:
\begin{equation}\label{eq:phi_1_erg_nonmix} \Phi_1 (\hat{U}) = W \left( \begin{array}{ccc} \sin(2J^z) \cos\bar{\theta} & \sin(2J^z) \sin\bar{\theta} & 0 \\ \sin(2J^z) \sin\bar{\theta} & -\sin(2J^z) \cos\bar{\theta} & 0 \\ 0 & 0 & -1 \end{array} \right) W^T , \end{equation} where $\bar{\theta} = \theta_Q + \theta_R$. The eigenvalues of $\Phi_1(\hat{U})$ are easily computed as:
\begin{eqnarray} z_1 &=& \quad -1, \\ z_2 &=& \sin(2J^z),  \\ z_3 &=& -\sin(2J^z) \end{eqnarray} and the corresponding eigenvectors, expressed as operators, are: \begin{eqnarray} \hat{A}_1 &=& \hat{w} \hat{\sigma}^z \hat{w}^\dagger , \\ \hat{A}_2 &=&  \cos\frac{\bar{\theta}}{2} \hat{w} \hat{\sigma}^x \hat{w}^\dagger + \sin \frac{\bar{\theta}}{2} \hat{w}\hat{\sigma}^y \hat{w}^\dagger , \\ \hat{A}_3 &=& -\sin\frac{\bar{\theta}}{2} \hat{w} \hat{\sigma}^x \hat{w}^\dagger + \cos \frac{\bar{\theta}}{2} \hat{w} \hat{\sigma}^y \hat{w}^\dagger . \end{eqnarray}
We have one peripheral eigenvalue $z_1 = -1$, and the other two eigenvalues $z_2$, $z_3$ strictly inside the unit circle (if $|J^z| \neq \pi/4$), as we set out to achieve [these eigenvalues are plotted in Fig. \ref{fig:lossless_remote_sensing}(a)].

We now discuss the implications for information transfer, focussing on the case of remote sensing of a continuous parameter $\lambda$. If the unknown parameter $\lambda$ is encoded in the initial state as:
\begin{equation} \hat{\rho}_{{\rm in},1}^{(\lambda)}(0) = \frac{1}{2} \left[ \hat{\mathbb{I}}_2 + f^{(\lambda)} \hat{A}_1 \right], \end{equation}
then after $t = N$ time steps the final spin state becomes:
\begin{eqnarray}
\hat{\rho}_{{\rm out}, 1}^{(\lambda)}(t) &\stackrel{(t=N)}{=}& \Phi_1^t [\hat{\rho}_{{\rm in},1}^{(\lambda)}(0)] \nonumber \\ &=& \frac{1}{2} \left[ \hat{\mathbb{I}}_2 + (-1)^{t} f^{(\lambda)} \hat{A}_1 \right],
\end{eqnarray} leading to lossless transmission of the encoded parameter information, by Result \ref{result:classes}\ref{item:lossless}. By contrast, if the parameter is encoded using the other eigenoperators:
\begin{equation} \hat{\rho}_{{\rm in},1}^{(\lambda)}(0) = \frac{1}{2} \left[ \hat{\mathbb{I}}_2 + f_2^{(\lambda)} \hat{A}_2 + f_3^{(\lambda)} \hat{A}_3 \right] , \label{eq:eta_F_loss} \end{equation} then the final state becomes:
\begin{eqnarray} \hat{\rho}_{{\rm out},1}^{(\lambda)}(t)  &\stackrel{(t=N)}{=}& \Phi_1^t [\hat{\rho}_{{\rm in},1}^{(\lambda)}(0)] \nonumber \\ &=& \frac{1}{2} \left[ \hat{\mathbb{I}}_2 + \sin^{t}(2J^z) \left( f_2^{(\lambda)} \hat{A}_2 + (-1)^{t} f_3^{(\lambda)} \hat{A}_3 \right) \right], \nonumber \end{eqnarray}
and transmitted information about $\lambda$ is exponentially suppressed with time $t=N$ (if $|J^z| \neq \pi/4$). Numerical simulations confirming both behaviours are shown in Fig.~\ref{fig:lossless_remote_sensing}(b).
\begin{figure}[b]
    \includegraphics[width=0.8\linewidth]{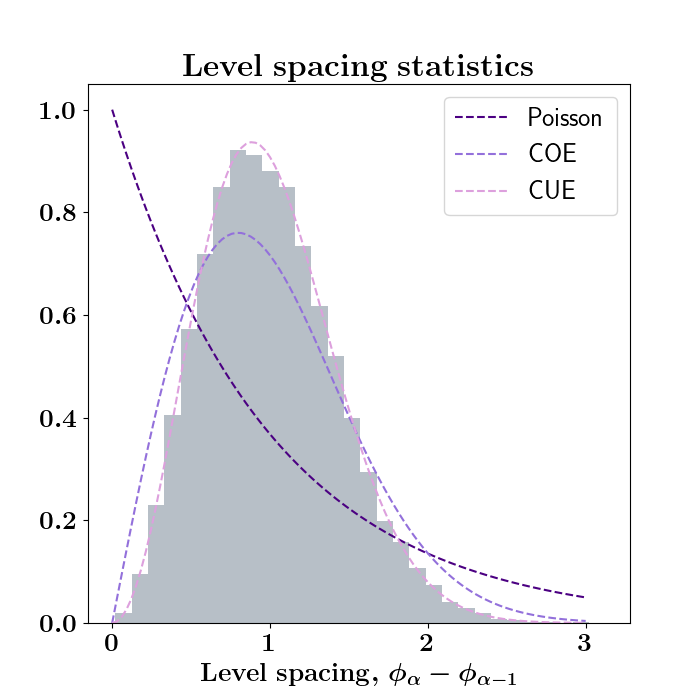}
    \caption{The level spacing statistics of the global Floquet unitary $\hat{\mathbb{U}}$, built with the unitary gate $\hat{U}$ constructed in Sec. \ref{sec:example}. The distribution of level spacings matches that of the circular unitary ensemble (CUE), indicating quantum chaos. Yet, the quantum channel $\Phi_1 (\hat{U})$ exhibits a peripheral eigenvalues indicating that the circuit supports lossless transmission of information.}
    \label{fig:levelspacing_erg_nonmix}
\end{figure}
To examine the global spectral properties of this model, we plot the level spacing statistics of the eigenphases $\phi_\alpha$ of the Floquet unitary $\hat{\mathbb{U}} \ket{\phi_\alpha} = e^{i\phi_\alpha} \ket{\phi_\alpha}$ in Fig.~\ref{fig:levelspacing_erg_nonmix}. The distribution agrees with the circular unitary ensemble (CUE), indicating quantum chaos. This is consistent with the results of Ref. \cite{Ber-21a}, which rigorously proved (by computing the spectral form factor in the long chain limit) that similar dual-unitary circuits are indeed quantum chaotic. Moreover, Ref. \cite{Ber-19a} developed an ergodic classification of dual-unitary circuits based on the dynamics of correlation functions, which in turn, are determined by the number of eigenvalues of the quantum channel $\Phi$ with unit value or unit modulus. In terms of that classification, the circuit described here is in the ergodic and non-mixing class, since it has an eigenvalue of unit modulus, but no non-trivial unit eigenvalues. 

The possibility of lossless remote sensing through a chaotic spin chain is particularly intriguing. Prior studies \cite{Doo-21a, Doo-23a, Doo-25a} have shown that similar sensing performance can be achieved in chaotic systems exhibiting quantum many-body scars (atypical non-thermal eigenstates in an otherwise thermal spectrum). However, in the model considered here, we find no evidence of scarring (see Appendix \ref{app:e}, where we plot the Floquet eigenstate entanglement entropy). Also, in Fig.~\ref{fig:lossless_remote_sensing}(c), we show that the Floquet eigenstate expectation values $\langle \phi_\alpha | \hat{\sigma}_N^\mu | \phi_\alpha \rangle$ are sharply concentrated around their thermal value $\langle \hat{\sigma}_N^\mu \rangle_{\rm thermal} = 0$, and increasingly concentrated with increasing system size $N$. This behaviour suggests that the diagonal ETH is satisfied, and that long time average of any qubit observable on the last site of the chain approaches the thermal value. 

Taking all of the above together, this example demonstrates that: 
\begin{result}
Lossless information transfer is possible through a chain of qubits that is chaotic, ergodic, and obeys the ETH.    
\end{result}

While it might seem paradoxical that a thermalising, non-integrable quantum chain can support perfect, lossless information transfer, the resolution lies in a subtle but crucial question of the order of limits. The system in question obeys the eigenstate thermalisation hypothesis, meaning that if takes the long-time limit $t \to \infty$ before the thermodynamic limit $N \to \infty$, the system will indeed thermalise and any initially localised information will be scrambled beyond recovery. However, if one takes the thermodynamic limit first, the local information will propagate through the chain perfectly along the lightcone and will not reach the boundary of the chain in any finite time. In other words, the two limits — long chain and long time — do not commute.

Another way to see this is to observe that, although information propagates without loss along the lightcone of the chain up to the time $t=N$, it is degraded after reflection at the boundary of the open chain at times $t > N$. Crucially, in our setup, the measurement of the last qudit takes place at the time $t=N$, before any reflection at the boundary of the open chain. Hence, perfect information transfer at an arbitrarily large time $t = N$ is consistent with ultimate thermalisation at longer times $t \gg N$.

\section{Conclusion}

In this work, we have investigated the transfer of information along the causal lightcone of a brickwork quantum circuit, focusing on a one-dimensional chain of qudits where information encoded in the first $M$ qudits is recovered by measurements on the last $M$ qudits. By restricting attention to the lightcone $t = N$, we showed that the problem reduces to the repeated application of a quantum channel $\Phi_M$, whose spectral properties completely determine whether information transfer is lossless or lossy. Specifically, lossless information transfer along the lightcone is possible if and only if $\Phi_M$ possesses peripheral eigenvalues (eigenvalues of unit modulus) while in their absence information decays irreversibly. For generic brickwork circuits, with two-qudit gates drawn from the circular unitary ensemble, peripheral eigenvalues are almost always absent, consistent with the non-integrable, thermalising character of such circuits.

Guided by the peripheral eigenvalue criterion, we identified classes of circuits that do support lossless transfer. For qubit chains ($d=2$) with $M = 1$, dual-unitarity is a necessary condition: peripheral eigenvalues of $\Phi_1 (\hat{U})$ can only arise when the two-qubit gate $\hat{U}$ is dual-unitary. However, this stringent requirement is relaxed in two natural extensions. For qubit chains with $M = 2$, we demonstrated numerically that non-dual-unitary gates $\hat{U}$ can also give rise to peripheral eigenvalues of $\Phi_2 (\hat{U})$, and proposed a conjecture characterising parameters of $\hat{U}$ for which this occurs. Similarly, for qutrit chains ($d = 3$) with $M = 1$, we found numerically that peripheral eigenvalues of $\Phi_1 (\hat{U})$ can appear without dual-unitarity of the two-qutrit gate $\hat{U}$. In both cases, our results suggest that a minimum — but not necessarily maximal — degree of operator entanglement in the gate $\hat{U}$ is required for peripheral eigenvalues and lossless transfer.
Interestingly, we find an explicit construction, in Sec.~\ref{sec:example}, of a dual-unitary brickwork circuit that supports lossless information transfer despite being quantum chaotic, ergodic, and satisfying the ETH. 


Our results open several directions for future investigation. On the analytical side, a complete characterisation of the conditions under which $\Phi_2$ (or more generally $\Phi_M$ for $M > 1$) admits peripheral eigenvalues remains an open problem, and a proof or disproof of Conjecture \ref{conj:M2_d2} would be a step forward. More broadly, it would be interesting to use the channel $\Phi_M$ to understand whether and how lossless information transfer can persist for smaller subsystems (e.g., a single qubit) inside the lighcone at times $t \neq N$. On the applied side, our results suggest that, in principle, even chaotic brickwork circuits can be viable platforms for quantum state transfer and remote quantum sensing, and it would be valuable to assess the robustness of the lossless transfer mechanisms identified here to noise, imperfect gates, and finite-size effects. More generally, this work highlights the power of minimal, exactly solvable circuit models as a framework for obtaining rigorous insights into quantum information dynamics in strongly interacting, nonintegrable systems.

\begin{acknowledgments}
This publication has emanated from research conducted with the financial support of Taighde \'{E}ireann – Research Ireland under Grant number 22/PATH-S/10812. We would like to thank Zolt\'{a}n Zimbor\'{a}s, Sergey N. Filippov, Nathan Keenan and Alessandro Summer for inspiring discussions in the early stages of this project. 
\end{acknowledgments}



\bibliography{refs}



\clearpage
\newpage

\onecolumngrid
\appendix
\section{The definition of $\Lambda_M^{\text{in/out}}$}

The $M-$qudit unitary matrix $\Lambda_M^{\rm in/out}$ only appears for $M\ge3$. For $M=3$, we define $\Lambda_3^{\rm in} = \mathcal{U}_2 \otimes \mathcal{I}$.
We can build the bigger unitaries, $\Lambda_4^{\rm in}~, \Lambda_5^{\rm in}~,\dots$, on top of $\Lambda_3^{\rm in}$ with the following recursive relations:
\begin{align}
    \Lambda_{M+1}^{\rm in} \stackrel{(M~\rm{ odd})}{=}& 
    \left(\Lambda_M \otimes \mathcal{I} \right) \left( \mathcal{I} \otimes \mathcal{U}_2^{\otimes \frac{M-1}{2}} \otimes \mathcal{I} \right) \\
    \Lambda_{M+1}^{\rm in} \stackrel{(M~\rm{ even})}{=}& \left(\Lambda_M \otimes \mathcal{I} \right)\left( \mathcal{U}_2^{\otimes \frac{M}{2}} \otimes \mathcal{I} \right).
\end{align}
Similarly, we define $\Lambda_3^{\rm out}  = \mathcal{I}\otimes\mathcal{U}_2$. And we can define $\Lambda_4^{\rm out},~\Lambda_5^{\rm out},\dots$ with the following relations:
\begin{align}
    \Lambda_{M+1}^{\rm out} \stackrel{(M,\rm{ odd})}{=}& 
     \left( \mathcal{I} \otimes \mathcal{U}_2^{\otimes \frac{M-1}{2}} \otimes \mathcal{I} \right) \left( \mathcal{I}\otimes \Lambda_M  \right) \\
    \Lambda_{M+1}^{\rm out} \stackrel{(M,\rm{ even})}{=}& \left( \mathcal{U}_2^{\otimes \frac{M}{2}}\otimes  \mathcal{I} \right) \left( \mathcal{I} \otimes \Lambda_M \right).
\end{align}

\label{app:a}

\section{Properties of eigenoperators of $\Phi_M$}

First, we demonstrate that the eigenoperators $\hat{A}_\mu$ of $\Phi_M$ can always be chosen to be traceless. We use the fact that the map $\Phi_M$ is a trace-preserving operation. That is,

\begin{align}
    \text{Tr}[\Phi_M(\hat{A})] =&  z_\mu \text{Tr}[\hat{A}] \\
    (1 -  z_\mu) \text{Tr}[\hat{A}] &= 0
\end{align}
The above equation is satisfied only if the eigenoperator $\hat{A}_\mu$ is a traceless operator, or if $z_\mu = 1$. However, the eigenoperators obeying $z_\mu = 1$ can also be chosen to be traceless. To see this, we observe that an operator with the eigenvalue $z_\mu = 1$ is always the trivial eigenoperator $\Phi_M [\hat{\mathbb{I}}] = \hat{\mathbb{I}}$. If there is an additional eigenoperator $\tilde{A}$ with eigenvalue $z_\mu = 1$ (i.e, the eigenvalue $1$ is degenerate), then any linear combination $a\hat{\mathbb{I}} + b\tilde{A}_\mu$ is also an eigenoperator obeying $\Phi_M [a\hat{\mathbb{I}} + b\tilde{A}_\mu] = a\hat{\mathbb{I}} + b\tilde{A}_\mu$, for arbitrary real coefficients $a$, $b$. These coefficients can always be chosen so that $a/b = - {\rm Tr}(\tilde{A})/{\rm Tr}(\hat{\mathbb{I}})$, ensuring that the eigenoperator $\hat{A}_\mu = a\mathbb{I} + b\tilde{A}_\mu$ is traceless.


Next, we show that for any operator $\hat{A}_\mu$ with a complex eigenvalue $z_\mu$, we also have an eigenoperator $\hat{A}_\mu^\dagger$ with eigenvalue $z_\mu^*$. The eigenvalue equation for $\hat{A}_\mu$ is:
\begin{equation}
    \Phi_M [\hat{A}_\mu] = \frac{1}{d}\text{Tr}_{q_1} \{ \mathcal{U}_{M+1}[ \hat{A}_\mu \otimes \mathbb{I} ] \} = z_\mu \hat{A}_\mu.
\end{equation}
Taking the Hermitian conjugate of this equation, we have:
\begin{equation}
    \Phi_M[\hat{A}^\dagger_\mu] =\frac{1}{d} \text{Tr}_{q_1} \{ \mathcal{U}^\dagger_{M+1}[ \hat{A}^\dagger_\mu \otimes \mathbb{I} ] \} = \frac{1}{d}\text{Tr}_{q_1} \{ \mathcal{U}_{M+1}[ \hat{A}^\dagger_\mu \otimes\mathbb{I} ] \} = z^{*}_\mu \hat{A}^\dagger_\mu,
\end{equation}
since $\mathcal{U}_M$ represents a unitary operation $\mathcal{U}_M = \mathcal{U}_M^\dagger$. Therefore, if $\hat{A}_\mu$ is an eigenoperator with complex eigenvalue $z_\mu$, then $\hat{A}^\dagger$ is also an eigenoperator with eigenvalue $z_\mu^{*}$.

\label{app:b}

\section{Lossless transmission of QFI via peripheral eigenvalues}
We calculate the scaling of the QFI under the dynamics of the quantum channel $\Phi_M$. First, we demonstrate the decay of $\eta_F$ under the channel $\Phi_M$ without peripheral eigenvalues, and then we show how $\Phi_M$ with a peripheral eigenvalue can preserve QFI.

Consider a quantum state $\hat{\rho}^{(\lambda)}_{{\rm in},M} (0)$, similar to Eq.~\ref{eq:rho_0_A_decomp}, encoded with a parameter $\lambda$,
\begin{equation} \hat{\rho}^{(\lambda)}_{{\rm in},M} (0) = \frac{1}{d^M} \Big\{ \hat{\mathbb{I}}_d^{\otimes M} + \sum_\mu [ f^{(\lambda)}_\mu \hat{A}_\mu + f_\mu^{(\lambda)*} \hat{A}_\mu^\dagger ] \Big\} ,
\end{equation}
where the complex coefficients $f_\mu^{(\lambda)}$ are functions of the parameter $\lambda$. The time-evolved reduced density matrix along the lightcone is:
\begin{align}
    \hat{\rho}^{(\lambda)}_{{\rm out},M} (t) &\stackrel{(t=N)}{=} \Phi_M^{N-M+1}[\hat{\rho}^{(\lambda)}_{{\rm in},M} (0)]\nonumber \\
    &= \frac{1}{d^M} \Big\{ \hat{\mathbb{I}}_d^{\otimes M} + \sum_\mu [ z^{N-M+1}_{\mu}f^{(\lambda)}_\mu \hat{A}_\mu + {\rm ~h.c.} ] \Big\} \\
    &= \frac{1}{d^M} \Big\{ \hat{\mathbb{I}}_d^{\otimes M} + \sum_\mu |z_\mu|^{N-M+1} \tilde{A}_\mu^{(\lambda)}(t) ] \Big\}.
\end{align}
where we have defined $\tilde{A}^{(\lambda)}_\mu = (e^{i\phi_\mu})^{N-M+1}f^{(\lambda)}_{\rm \mu} \hat{A}_{\rm \mu} + {\rm ~h.c.} $ and used the representation $z_\mu = |z_\mu|e^{i \phi_\mu}$.
The operator $\tilde{A}^{(\lambda)}_\mu$, is an oscillatory term, and $|z_{\rm \mu}|^{N-M+1}$ is the dissipation on $\tilde{A}_\mu^{(\lambda)}$.

If the maximum eigenvalue $|z_{\rm max}| < 1$, only $\tilde{A}_{\rm max}^{(\lambda)}$ dominates in the long time limit, and the time-evolved density matrix $\hat{\rho}^{(\lambda)}_{{\rm out},M} (t)$ can be approximated to 
\begin{equation}
    \hat{\rho}^{(\lambda)}_{{\rm out},M} (t) \approx \frac{1}{d^M} \Big\{ \hat{\mathbb{I}}_d^{\otimes M} + |z_{\rm max}|^{N-M+1}\tilde{A}^{(\lambda)}_{\rm max}(t) \Big\}.
\end{equation}
The operator $\tilde{A}^{(\lambda)}_{\rm max}(t)$ is a Hermitian operator and can be diagonalized to $\tilde{A}^{(\lambda)}_{\rm max}(t) = \sum_i a_i\ket{a_i}\bra{a_i}$, where $\ket{a_i}$ serves as an eigenvector for the approximated density matrix $\hat{\rho}^{(\lambda)}_{{\rm out},M} (t)$, with the eigenvalues $p_i\approx (1 + |z_{\rm max}|^{N-M+1}a_i)/d^M$. 
The derivative of the time-evolved state can be approximated to 
\begin{equation}
    \partial_\lambda\hat{\rho}^{(\lambda)}_{{\rm out},M} (t) \approx \frac{1}{d^M} \Big\{ |z_{\rm max}|^{N-M+1}\partial_\lambda\tilde{A}^{(\lambda)}_{\rm max}(t) \Big\}
\end{equation}
Using the derivative $\partial_\lambda\hat{\rho}^{(\lambda)}_{{\rm out},M} (t)$ and the probability $p_i$ in Eq.~\ref{eq:QFI}, we arrive at the following expression for QFI,
\begin{equation}
    F_Q[\hat{\rho}^{(\lambda)}_{{\rm out},M}(t)] \approx |z_{\rm max}|^{N-M+1}\sum_{ij} \frac{ |  \bra{a_i}\partial_\lambda \tilde{A}_{\rm max}^{(\lambda)}(t)\ket{a_j}|^2}{[2+|z_{\rm max}|^{N-M+1}(a_i + a_j)]/d^M}.
\end{equation}
The numerator $\bra{a_i}\partial_\lambda \tilde{A}_{\rm max}^{(\lambda)}(t)\ket{a_j}$ is an oscillatory term and is not expected to contribute to the decay in the long time limit. 
The denominator converges to $2/d^M$ in the long-time limit. Therefore, QFI is dominated by the leading eigenvalue scaling $|z_{\rm max}|^{N-M+1}$.
When the QFI is compared with the initial QFI, Eq.~\ref{eq:etaF}, we arrive at the scaling 
\begin{equation}
\eta_F \sim |z_{\rm max}|^{N-M+1}.
\end{equation}

If the map $\Phi_M$ has peripheral eigenvalues $z_\mu = e^{i \phi}$, we consider the initial state restricted to the periphery.
Then, the time-evolved state is given by 
\begin{align}
    \hat{\rho}^{(\lambda)}_{{\rm out},M} (t) &\stackrel{(t=N)}{=} \Phi_M^{N-M+1}[\hat{\rho}^{(\lambda)}_{{\rm in},M} (0)]\nonumber \\
    &= \frac{1}{d^M} \Big\{ \hat{\mathbb{I}}_d^{\otimes M} + \sum_\mu \tilde{A}_\mu^{(\lambda)}(t) \Big\}
\end{align}
where $\tilde{A}^{(\lambda)}_\mu(t) = (e^{i\phi_\mu})^{N-M+1}f^{(\lambda)}_{\rm \mu} \hat{A}_{\rm \mu} + {\rm ~h.c.} $
The term $\sum_\mu \tilde{A}_\mu^{(\lambda)}(t)$, is an oscillatory term with no dissipation, which is expected to return to its initial form at some time $t^*$, due to the Poincar\'{e} recurrence theorem. However, the QFI cannot increase under the operation $\Phi_M$, as we have $F_Q[\Phi[\hat{\rho}^{(\lambda)}_{{\rm out},M} (t)] \le F_Q[\hat{\rho}^{(\lambda)}_{{\rm out},M} (t)]$, and therefore, the QFI content of the time-evolved quantum state must remain the same under the evolution through the map $\Phi_M$.

\label{app:c}

\section{Singular Values of $\Phi_1$}
We calculate the singular values of $\Phi_1$ defined in Eq.~\ref{eq:Phi_1^XX}-\ref{eq:Phi_1^ZZ}.
The matrix elements $\Phi_{\xi\xi'}$ in Eq.~\ref{eq:Phi_matrix_elements_SVD} are given by 
\begin{equation}\label{eq:app_svd_cal}
    \Phi_1^{\xi \xi'}(\hat{U}) = \frac{1}{4} {\rm Tr} [(\hat{\mathbb{I}}_2 \otimes \hat{\sigma}^{\xi'} ) \hat{V} (\hat{\sigma}^\xi  \otimes \hat{\mathbb{I}}_2 ) \hat{V}^\dagger],
\end{equation}
where $\hat{V} = e^{-i(J^x \hat{\sigma}^x \otimes\hat{\sigma}^x + J^y \hat{\sigma}^y \otimes\hat{\sigma}^y + J^x \hat{\sigma}^z \otimes\hat{\sigma}^z)} = e^{-iJ^x \hat{\sigma}^x \otimes\hat{\sigma}^x}e^{-iJ^y\hat{\sigma}^y \otimes\hat{\sigma}^y}e^{-iJ^z\hat{\sigma}^z\otimes\hat{\sigma}^z}$, since $[\hat{\sigma}^\alpha \otimes \hat{\sigma}^\alpha, \hat{\sigma}^\beta \otimes \hat{\sigma}^\beta]=0$.
We will show the calculation for $\Phi_1^{zz}$, and a similar calculations can be performed for $\Phi_1^{xx}$ and $\Phi_1^{yy}$. To calculate the matrix element $\Phi_1^{xx}$, we will use the following identities:
\begin{align}\label{eq:iden1}
    e^{-iJ^y \hat{\sigma}^y \otimes \hat{\sigma}^y}(\hat{\sigma}^z \otimes \hat{\mathbb{I}}_2) e^{iJ^y \hat{\sigma}^y \otimes \hat{\sigma}^y}&= \left[ \cos J^y (\hat{\mathbb{I}}_2\otimes\hat{\mathbb{I}}_2) - i\sin J^y (\hat{\sigma}^y \otimes \hat{\sigma}^y) \right] \nonumber \\  &(\hat{\sigma}^z \otimes \hat{\mathbb{I}}_2) \left[ \cos J^y (\hat{\mathbb{I}}_2\otimes\hat{\mathbb{I}}_2) + i\sin J^y (\hat{\sigma}^y \otimes \hat{\sigma}^y) \right] \nonumber \\
    & = \cos2J^y (\hat{\sigma}^z \otimes \hat{\mathbb{I}}_2) - \sin2J^y (\hat{\sigma}^x \otimes \hat{\sigma}^y ),
\end{align}

\begin{align}\label{eq:iden2}
    e^{-iJ^x \hat{\sigma}^x \otimes \hat{\sigma}^x}(\hat{\sigma}^z \otimes \hat{\mathbb{I}}_2) e^{iJ^x \hat{\sigma}^x \otimes \hat{\sigma}^x}&= \left[ \cos J^x (\hat{\mathbb{I}}_2\otimes\hat{\mathbb{I}}_2) - i\sin J^x (\hat{\sigma}^x \otimes \hat{\sigma}^x) \right] \nonumber \\  &(\hat{\sigma}^z \otimes \hat{\mathbb{I}}_2) \left[ \cos J^x (\hat{\mathbb{I}}_2\otimes\hat{\mathbb{I}}_2) + i\sin J^x (\hat{\sigma}^x \otimes \hat{\sigma}^x) \right] \nonumber \\
    & = \cos2J^x (\hat{\sigma}^z \otimes \hat{\mathbb{I}}_2) + \sin2J^x (\hat{\sigma}^x \otimes \hat{\sigma}^y ), 
\end{align}
and 
\begin{align}\label{eq:iden3}
    e^{-iJ^x \hat{\sigma}^x \otimes \hat{\sigma}^x}(\hat{\sigma}^x \otimes \hat{\sigma}^y) e^{iJ^x \hat{\sigma}^x \otimes \hat{\sigma}^x}&= \left[ \cos J^x (\hat{\mathbb{I}}_2\otimes\hat{\mathbb{I}}_2) - i\sin J^x (\hat{\sigma}^x \otimes \hat{\sigma}^x) \right] \nonumber \
    \\  &(\hat{\sigma}^x \otimes \hat{\sigma}^y) \left[ \cos J^x (\hat{\mathbb{I}}_2\otimes\hat{\mathbb{I}}_2) + i\sin J^x (\hat{\sigma}^x \otimes \hat{\sigma}^x) \right] \nonumber \\
    & = \cos2J^x (\hat{\sigma}^x \otimes \hat{\sigma}^y) + \sin2J^x (\hat{\mathbb{I}}_2 \otimes \hat{\sigma}^z ).
\end{align}
The evolution $\hat{V}(\hat{\sigma}^z \otimes \hat{\mathbb{I}}_2)\hat{V}^\dagger$ in Eq.~\ref{eq:app_svd_cal}, now becomes 
\begin{align}
    \hat{V}(\hat{\sigma}^z \otimes \hat{\mathbb{I}}_2)\hat{V}^\dagger &= e^{-iJ^x \hat{\sigma}^x \otimes\hat{\sigma}^x}e^{-iJ^y\hat{\sigma}^y \otimes\hat{\sigma}^y}e^{-iJ^z\hat{\sigma}^z\otimes\hat{\sigma}^z}  (\hat{\sigma}^z \otimes \hat{\mathbb{I}}_2)e^{iJ^z \hat{\sigma}^z \otimes\hat{\sigma}^z}e^{-iJ^y\hat{\sigma}^y \otimes\hat{\sigma}^y}e^{-iJ^x\hat{\sigma}^x\otimes\hat{\sigma}^x}\nonumber \\
    & = e^{-iJ^x \hat{\sigma}^x \otimes\hat{\sigma}^x}e^{-iJ^y\hat{\sigma}^y \otimes\hat{\sigma}^y}(\hat{\sigma}^z \otimes \hat{\mathbb{I}}_2)e^{iJ^y\hat{\sigma}^y \otimes\hat{\sigma}^y}e^{iJ^x\hat{\sigma}^x\otimes\hat{\sigma}^x}.
\end{align}
Using Eq.~\ref{eq:iden1}, we have,
\begin{equation}
    \hat{V}(\hat{\sigma}^z \otimes \hat{\mathbb{I}}_2)\hat{V}^\dagger =e^{-iJ^x \hat{\sigma}^x \otimes\hat{\sigma}^x}\left[ \cos2J^y (\hat{\sigma}^z \otimes \hat{\mathbb{I}}_2) - \sin2J^y (\hat{\sigma}^x \otimes \hat{\sigma}^y ) \right]e^{iJ^x \hat{\sigma}^x \otimes\hat{\sigma}^x}.
\end{equation}
Now, using Eq.~\ref{eq:iden2} and Eq.~\ref{eq:iden3}, we arrive at 
\begin{align}
\hat{V}(\hat{\sigma}^z \otimes \hat{\mathbb{I}}_2)\hat{V}^\dagger = & \cos2J^x\cos2J^y (\hat{\sigma}^z \otimes \hat{\mathbb{I}}_2) - \sin2J^x\cos2J^y(\hat{\sigma}^y\otimes\hat{\sigma}^x)\nonumber \\
    &+\sin2J^y\cos2J^x(\hat{\sigma}^x\otimes\hat{\sigma}^y) + \sin2J^x\sin2J^y(\hat{\mathbb{I}}_2 \otimes \hat{\sigma}^z).
\end{align}
Substituting this in Eq.~\ref{eq:app_svd_cal}, we calculate
$\Phi_1^{zx}= \Phi_1^{zy}=0$, and $\Phi_1^{zz} = \sin2J^x\sin2J^y$.

\label{app:d}

\section{Half-Chain entanglement entropy of dual-unitary with peripheral eigenvalues}

\begin{figure*}[t]
    \includegraphics[width=0.5\linewidth]{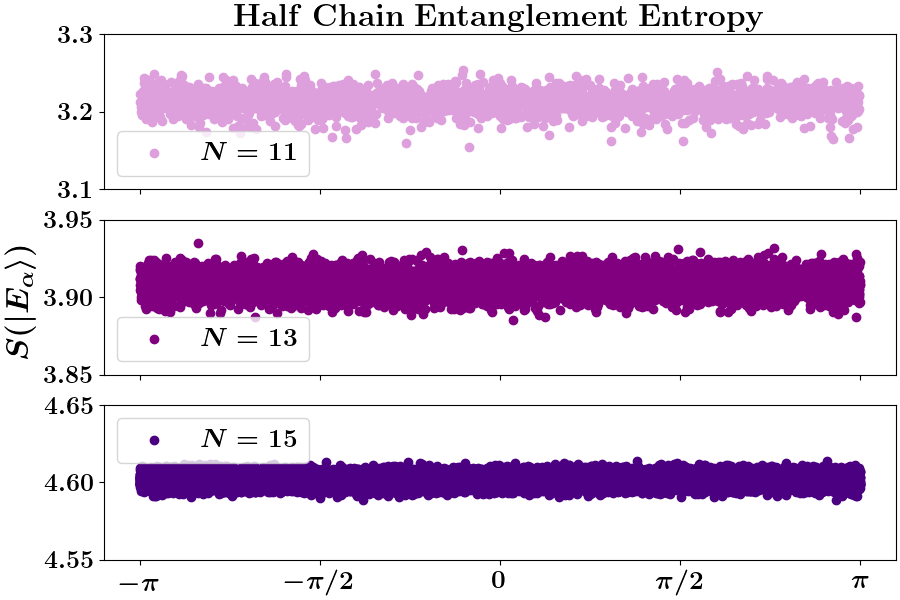}
    \caption{The half-chain entanglement entropy for the eigenstates $\ket{\phi_\alpha}$ of the Floquet operator $\mathbb{U}$ that supports lossless transmission of QFI along the chain of qubits. The entanglement entropy does not show any characteristics of quantum many-body scars.}
    \label{fig:ee_entropy}
\end{figure*}

Quantum many-body scars serve as an example of eigenstates of a non-integrable Hamiltonian that violate ETH.
Lossless transmission of QFI along the chain of qubits opens up the question of the presence of scars in the system. Half-chain entanglement entropy of the eigenstates of the Floquet operator $\mathbb{U}$ can reveal the presence of scars in many-body systems, as scars typically have low entanglement entropy as compared to the eigenstates that thermalize. The half-chain entanglement entropy of the eigenstates $\ket{\phi_\alpha}$ of the Floquet operator $\mathbb{U}$, which supports lossless transmission of QFI for $M=1$, shows no signatures of quantum scarring, see Fig~\ref{fig:ee_entropy}. 

\label{app:e}

\end{document}